\pdfoutput=1
\documentclass[twocolumn,aps,preprintnumbers,amsmath,amssymb,nofootinbib]{revtex4-1}

\usepackage{epsfig}
\usepackage{amsmath}
\usepackage{amssymb}
\usepackage{subfigure}
\usepackage{cancel}
\usepackage{textcomp}
\usepackage{calc}
\usepackage{graphicx}
\usepackage{dcolumn}
\usepackage{color}
\usepackage{xcolor}
\usepackage{pifont}
\usepackage{slashed}
\usepackage{bbold}
\usepackage{xfrac}

\usepackage{hyperref}
\hypersetup{colorlinks=true,urlcolor=blue,linkcolor=red,citecolor=green!60!black}

\begin{document}

%%%%%%%%%%%%%%%%%%%%%%%%%%%%%%%%%%%%%%%%%%%%%%%%%%%%%%%%%%%%%%%%%%%%%%%%%%%%%%%%%%%%%%%%%%%%%%%%%%%%

\preprint{MS-TP-24-11}

\title{Gravitational waves from low-scale cosmic strings}

\author{Kai Schmitz}
\email{kai.schmitz@uni-muenster.de}

\author{Tobias Schr\"oder}
\email{schroeder.tobias@uni-muenster.de}

\affiliation{Institute for Theoretical Physics, University of M\"unster, 48149 M\"unster, Germany}

%%%%%%%%%%%%%%%%%%%%%%%%%%%%%%%%%%%%%%%%%%%%%%%%%%%%%%%%%%%%%%%%%%%%%%%%%%%%%%%%%%%%%%%%%%%%%%%%%%%%

\begin{abstract}
Cosmic strings are a common prediction in many grand unified theories and a promising source of stochastic gravitational waves (GWs) from the early Universe. In this paper, we point out that the GW signal from cosmic strings produced at a comparatively low energy scale, $v \lesssim 10^9\,\textrm{GeV}$, exhibits several novel features that are not present in the case of high-scale cosmic strings. Our findings notably include (i) a sharp cutoff frequency $f_{\rm cut}$ in the GW spectrum from the fundamental oscillation mode on closed string loops and (ii) an oscillating pattern in the total GW spectrum from all oscillation modes whose local minima are located at integer multiples of $f_{\rm cut}$. These features reflect the fact that string loops produced in the early Universe fail to shrink to zero size because of GW emission within the age of the Universe, if their tension is low enough. In addition, they offer an exciting opportunity to directly probe the discrete spectrum of oscillation modes on closed string loops in GW observations. For strings produced at a scale $v \sim 10^9\,\textrm{GeV}$, the novel features in the GW spectrum are within the sensitivity reach of future experiments such as BBO and DECIGO.
\end{abstract}

%%%%%%%%%%%%%%%%%%%%%%%%%%%%%%%%%%%%%%%%%%%%%%%%%%%%%%%%%%%%%%%%%%%%%%%%%%%%%%%%%%%%%%%%%%%%%%%%%%%%

\date{\today}
\maketitle

%%%%%%%%%%%%%%%%%%%%%%%%%%%%%%%%%%%%%%%%%%%%%%%%%%%%%%%%%%%%%%%%%%%%%%%%%%%%%%%%%%%%%%%%%%%%%%%%%%%%

\section{Introduction}
\label{sec:intro}

Cosmic strings~\cite{Vilenkin:1984ib,Hindmarsh:1994re,Vilenkin_Shellard_2000} are thin tube-like topological defects that can form during phase transitions in the early Universe via the Kibble--Zurek mechanism~\cite{Kibble:1976sj,Kibble:1980mv,Zurek:1985qw}. A prerequisite for the formation of a cosmic-string network is that the underlying phase transition must spontaneously break a symmetry of the theory and end on a vacuum manifold $\mathcal{M}$ with nontrivial first homotopy group, $\pi_1\left(\mathcal{M}\right) \ncong \left\{\mathbb{1}\right\}$. This condition is readily satisfied in many grand unified theories (GUTs)~\cite{Jeannerot:2003qv}, e.g., in intermediate GUT symmetry breaking steps that spontaneously break a local $U(1)$ symmetry. The energy scale of $U(1)$ symmetry breaking then controls the tension $\mu$ (i.e., energy per length) of the cosmic strings. Indeed, if the $U(1)$ symmetry is broken by the vacuum expectation value of a complex $U(1)$-charged scalar field, $v = \left<\phi\right>$, the resulting strings will have tension $\mu \sim 2\pi v^2$~\cite{Bogomolny:1975de}, or equivalently,
\begin{equation}
\label{eq:Gmuv}
G\mu \sim 4 \times 10^{-8} \left(\frac{v}{10^{15}\,\textrm{GeV}}\right)^2 \,,
\end{equation}
where $G = 1/\left(8\pi M_{\rm Pl}^2\right)$ is Newton's constant and $M_{\rm Pl} \simeq 2.435 \times 10^{18}\,\textrm{GeV}$ denotes the reduced Planck mass. 
$G\mu$ is a dimensionless measure of the string tension and of central importance for the phenomenology of cosmic strings.

In recent years, cosmic strings have received a considerable amount of attention, as they represent a promising source of gravitational waves (GWs) from the early Universe~\cite{Vilenkin:1981bx,Vachaspati:1984gt,Damour:2004kw} and thus constitute an attractive target for ongoing and upcoming GW observations~\cite{LIGOScientific:2017ikf,Auclair:2019wcv,LIGOScientific:2021nrg}. This notably includes pulsar timing array (PTA) searches for a stochastic GW background (GWB), which recently witnessed an important milestone: the first compelling evidence for a GWB signal at nanohertz frequencies~\cite{NANOGrav:2023gor,EPTA:2023fyk,Reardon:2023gzh,Xu:2023wog}. This signal, assuming it to be a genuine GWB signal, is believed to be of astrophysical origin and sourced by a cosmic population of inspiraling supermassive black-hole binaries (SMBHBs)~\cite{NANOGrav:2023hfp} and/or of cosmological origin and sourced by new particle physics in the early Universe~\cite{NANOGrav:2023hvm,EPTA:2023xxk}. Cosmic strings belong to the latter class of possible sources in the PTA band~\cite{Ellis:2020ena,Blasi:2020mfx,Blanco-Pillado:2021ygr}. As shown in Ref.~\cite{NANOGrav:2023hvm}, ordinary stable cosmic strings in the Nambu--Goto approximation can fit the signal in the NANOGrav 15-year data set for values of the string tension of around $G\mu \sim 3 \times 10^{-11}$. The quality of the fit is, however, poor, and stable strings as the sole explanation of the PTA signal are disfavored in comparison to all other models considered in Ref.~\cite{NANOGrav:2023hvm}. It is therefore reasonable to assume that the PTA signal must have a different origin, either in the form of SMBHBs or other exotic physics in the early Universe (e.g., metastable cosmic strings~\cite{Buchmuller:2020lbh,Buchmuller:2021mbb,Buchmuller:2023aus}), in which case one is able to derive an upper limit on the allowed tension of stable strings, $G\mu \lesssim 10^{-10}$~\cite{NANOGrav:2023hvm}. 

The upper bound on $G\mu$ from recent PTA observations needs to be contrasted with the sensitivity reach of future GW experiments like LISA, which promises to be sensitive to string tensions as small as $G\mu \sim 10^{-17}$~\cite{Auclair:2019wcv,Blanco-Pillado:2024aca}. Together, these two values span a range of seven orders of magnitude in the string tension that will be explored by the first generation of space-borne GW interferometers (LISA~\cite{LISA:2017pwj}, Taiji~\cite{Hu:2017mde}, and TianQin~\cite{TianQin:2015yph}). According to Eq.~\eqref{eq:Gmuv}, this range of $G\mu$ values translates to $U(1)$ symmetry breaking scales in the range roughly from $v \sim 10^{10}\,\textrm{GeV}$ to $v \sim 10^{14}\,\textrm{GeV}$. The GWB signal from a string network that is generated at such a high energy scale has been well studied in the literature~\cite{Blanco-Pillado:2017oxo,Gouttenoire:2019kij,Blanco-Pillado:2019tbi}. 

Meanwhile, it is conceivable that cosmic strings may also form at lower energies, $v \ll 10^{10}\,\textrm{GeV}$. In fact, the $U(1)$ symmetry responsible for the generation of cosmic strings may only be broken shortly above the electroweak scale, such that $G\mu \sim 10^{-33}$. Cosmic strings with such a low tension will hardly lead to any observational GWB signal; however, the situation is different if the string energy scale $v$ is low, $v \ll 10^{10}\,\textrm{GeV}$, but still large enough such that the resulting GWB signal is within the reach of second-generation space-borne GW interferometers (BBO~\cite{Corbin:2005ny} and DECIGO~\cite{Seto:2001qf}). As we will see, such a scenario requires string energy scales of around $v \sim 10^9\,\textrm{GeV}$, which may be easily realized in many extensions of the Standard Model. An obvious candidate for a $U(1)$ symmetry that can get spontaneously broken at $v \sim 10^9\,\textrm{GeV}$ is $U(1)_{B-L}$, the Abelian gauge symmetry associated with baryon-minus-lepton number $B\!-\!L$. On top of cosmic $B\!-\!L$ strings, the spontaneous breaking of $U(1)_{B-L}$ at $v \sim 10^9\,\textrm{GeV}$ might result in the generation of heavy Majorana masses for a set of sterile right-handed neutrinos of the order of $v$ (or lower), which would set the stage for the seesaw mechanism and the generation of the baryon asymmetry via leptogenesis~\cite{Buchmuller:2013lra,Dror:2019syi,Buchmuller:2019gfy}. But numerous other scenarios are possible, and in this work, we will refrain from specifying the microscopic model that gives rise to low-scale cosmic strings. 

Instead, the goal of the present paper is to point out that the GWB spectrum from low-scale cosmic strings can be qualitatively different from the well-known GWB spectrum from high-scale cosmic strings. This observation is based on a general and simple argument: The GWB signal from a cosmic-string network mostly arises from the GW emission of closed string loops~\cite{Vachaspati:1984gt,Damour:2001bk}, which continuously shrink as they emit GWs. The rate at which a string loop of length $\ell$ shrinks because of GW emission is, however, directly proportional to $G\mu$,
\begin{equation}
\label{eq:dldt}
\frac{d\ell}{dt} = - \Gamma G\mu \,,
\end{equation}
where $\Gamma \simeq 50$ follows from numerical simulations~\cite{Blanco-Pillado:2017oxo}. For a given initial loop length $\ell_{\rm ini}$, there is hence a minimal value of $G\mu$ below which energy loss via GW emission no longer suffices to drive string loops to zero size between their time of production and today. For typical values of $\ell_{\rm ini}$, we find that the relevant $G\mu$ threshold value can be as large as $G\mu \sim 10^{-19}$, corresponding to a string energy scale of roughly $v \sim 10^9\,\textrm{GeV}$. At $G\mu$ values below this threshold, no string loop produced in the early Universe has yet managed to fully disappear because of GW emission. Instead, the length of loops present in the string network is always bounded from below, even up to the present time, which implies the existence of a sharp ultraviolet (UV) cutoff frequency $f_{\rm cut}$ in the contribution to the GW spectrum that is sourced by the fundamental oscillation mode on closed string loops. In the total GW spectrum, which receives contributions from all loop oscillation modes, this cutoff frequency induces a series of characteristic peaks and dips, where the dips are located at integer multiples of $f_{\rm cut}$. These novel features in the GW spectrum from low-scale strings are unique, especially the integer-spaced sequence of dips at $f_{\rm cut}$, $2f_{\rm cut}$, etc., and not present in the GWB spectrum from high-scale strings. Moreover, as long as the string energy scale remains close to $v \sim 10^9\,\textrm{GeV}$, these features could possibly be observed in the future, which offers the exciting prospect to probe the discrete spectrum of oscillation modes on string loops in future GW observations.

The rest of the paper is organized as follows. In the next section, we will present our argument in more detail, derive an explicit expression for $f_{\rm cut}$, and discuss several possible choices for the smallest loop length $\ell_{\rm ini}$ at the time of formation in the early Universe. In Sec.~\ref{sec:vos}, we will then compute analytical expressions for the GWB spectrum from low-scale strings based on the velocity-dependent one-scale (VOS) model~\cite{Martins:1996jp,Martins:2000cs,Sousa:2013aaa}, which provides a simple and illustrative understanding of the novel features in the spectrum. The central assumption of the VOS model is that the string network is in the scaling regime and all relevant scales of the network are controlled by only one relevant length scale, e.g., the particle horizon $d_h$. The VOS model notably assumes that new loops in the network are always born with a length $\ell$ that is a fixed fraction of the horizon scale, $x = \ell/d_h = \alpha$, where, e.g., $\alpha \simeq 0.05$ deep in the radiation-dominated era~\cite{Blanco-Pillado:2013qja}. As we will demonstrate in Sec.~\ref{sec:vos}, this choice of initial loop length at the time of formation (essentially a delta function on the $\ln x$ axis) results in a series of sharp and pronounced peaks and dips in the GWB spectrum. Numerical simulations of string networks, however, show that the distribution of initial loop lengths does have a finite (albeit small) width~\cite{Blanco-Pillado:2013qja}. In Sec.~\ref{sec:bos}, we will therefore refine our analysis and account for the distribution of initial loops lengths found in the simulations by Blanco-Pillado, Olum, and Shlaer (BOS). To this end, we will first generalize the well-known BOS loop number density during radiation domination, replacing the delta function in the loop production function by a Gaussian distribution on the $\ln x$ axis, and then show how the finite width of this distribution results in a smoothing of the peaks and dips in the GWB spectrum. As a result, we will be able to conclude that, for a realistic distribution of loop lengths at the time of formation, the characteristic series of peaks and dips in the GWB spectrum from low-scale strings becomes slightly washed out but still remains at a detectable level. Sec.~\ref{sec:conclusions} finally contains our conclusions and a brief outlook. In particular, we will comment on how our analysis can be readily generalized from scaling models of the string network, such as the VOS and BOS models, which we consider in the present work, to nonscaling models such as the one studied in Ref.~\cite{Auclair:2019jip}, which explicitly takes into account the effect of particle radiation at the level of Eq.~\eqref{eq:dldt} and which we will consider in future work~\cite{Schmitz:2024abc}. We will discuss the robustness of our results and argue that we expect qualitatively and quantitatively  similar results for the model in Ref.~\cite{Auclair:2019jip}.

%%%%%%%%%%%%%%%%%%%%%%%%%%%%%%%%%%%%%%%%%%%%%%%%%%%%%%%%%%%%%%%%%%%%%%%%%%%%%%%%%%%%%%%%%%%%%%%%%%%%

\section{Sharp cutoff frequency}
\label{sec:cut}

The GWB signal from a string network is dominated by the GW emission from string loops. The long strings in the network, i.e., infinitely long strings or super-horizon loops, also emit GWs~\cite{Figueroa:2012kw,Figueroa:2020lvo}. However, their contribution to the total signal turns out to be strongly suppressed compared to the contributions from loops~\cite{Buchmuller:2013lra}; see Sec.~4.4 in Ref.~\cite{Auclair:2019wcv} for an extended discussion. For high-scale strings, GW emission from loops and long strings during radiation domination notably results in two flat, plateau-like contributions to the GWB spectrum whose amplitudes scales like $v$ and $v^4$, respectively. These scaling relations illustrate that the suppression of the long-string contribution to the GWB spectrum is more pronounced at smaller string energy scales $v$. In the following, we will therefore only discuss GWs from loops and ignore the GW signal from long strings.

\subsection{Principal argument}

String loops of length $\ell$ emit GWs at frequencies corresponding to their fundamental oscillation mode, $f = 2/\ell$, as well as their higher harmonic excitations, $f = 4/\ell, 6/\ell, 8/\ell, \cdots$. The expansion of the Universe between the time of emission and today causes in addition a redshift of these frequencies, such that GWs emitted by a loop of length $\ell\left(t\right)$ at time $t$ are observed today at 
\begin{equation}
\label{eq:flk}
f = \frac{a\left(t\right)}{a_0}\frac{2k}{\ell\left(t\right)} \,, \qquad k = 1,2,3,\cdots \,,
\end{equation}
where $a$ is the scale factor in the Friedmann--Lema\^itre--Robertson--Walker metric. Meanwhile, loops shrink as they emit GWs according to the decay law in Eq.~\eqref{eq:dldt}. Assuming an initial loop length $\ell_*$ at the time of formation $t_*$, the integral of Eq.~\eqref{eq:dldt} yields the linear relation
\begin{equation}
\label{eq:lt}
\ell\left(t\right) = \ell_* - \Gamma G\mu \left(t-t_*\right) \,.
\end{equation}
Numerical simulations show that, during radiation domination, the distribution of initial lengths $\ell_*$ is sharply peaked around a fixed fraction of the horizon scale~\cite{Blanco-Pillado:2013qja},
\begin{equation}
\label{eq:lstar}
\ell_* = \alpha\,d_h\left(t_*\right) \,, \qquad \alpha \simeq 0.05 \,, \qquad d_h\left(t_*\right) = 2t_* \,, 
\end{equation}
such that the size of the loops produced in the network steadily increases as the Universe expands and $d_h$ grows.

Let us now focus on the contribution to the GWB spectrum from fundamental string oscillations, i.e., the $k=1$ mode in Eq.~\eqref{eq:flk}, and assume that $G\mu$ is in fact so small that no loop ever produced has yet reached zero length,
\begin{equation}
\label{eq:lt0}
\ell\left(t_0\right) > 0 \quad\textrm{for all}\quad t_* \in \left[t_{\rm ini}, t_0\right] \,.
\end{equation}
From the combination of Eqs.~\eqref{eq:flk}, \eqref{eq:lt}, and \eqref{eq:lstar}, we are then able to conclude that the highest frequencies in the present-day GWB spectrum from fundamental string oscillations (i) correspond to GWs that have experienced the least amount of redshift and that (ii) arise from the shortest loops in the network. In other words, the highest frequencies belong to GWs that are emitted at present by the shortest loops. These loops correspond in turn to the smallest loops that were ever produced and that had the most time to shrink since their production, i.e., the loops that formed right at $t_{\rm ini}$, when the standard treatment of GW emission from a scaling network becomes applicable. We will discuss the precise definition and numerical value of $t_{\rm ini}$ shortly. For now, we are more interested in the loops that form at $t_{\rm ini}$ at length $\ell_*\left(t_{\rm ini}\right) = 2\alpha\,t_{\rm ini}$. Today, these loops have length
\begin{equation}
\ell_{\rm min} = \ell_*\left(t_{\rm ini}\right) - \Gamma G\mu\left(t_0 - t_{\rm ini}\right) \,, 
\end{equation}
and emit GWs at the UV cutoff frequency
\begin{equation}
\label{eq:fcut}
\boxed{f_{\rm cut} = \frac{2}{\ell_{\rm min}} = \frac{2}{2\alpha\,t_{\rm ini}-\Gamma G\mu\left(t_0 - t_{\rm ini}\right)} \,.}
\end{equation}

Indeed, as long as this frequency is positive and finite, $0 < f_{\rm cut} < \infty$, a sharp UV cutoff will be present in the GWB spectrum from fundamental loop oscillations. This is the case as long as $G\mu$ is small enough, such that
\begin{equation}
\Gamma G\mu\left(t_0 - t_{\rm ini}\right) < 2\alpha\,t_{\rm ini} \,,
\end{equation}
or equivalently, since $t_0 \ggg t_{\rm ini}$, 
\begin{equation}
\label{eq:tcut}
t_{\rm ini} > t_{\rm cut} = \frac{\Gamma G\mu}{2\alpha}\,t_0 \simeq 2.2\,\textrm{s}\left(\frac{G\mu}{10^{-20}}\right) \,,
\end{equation}
where we set $\Gamma = 50$ and $\alpha = 0.05$. For the values of $G\mu$ that we will be interested in, $t_{\rm cut}$ always falls into the radiation-dominated era. It is therefore convenient to express the condition $t_{\rm ini} > t_{\rm cut}$ in terms of the temperature of the thermal bath during the radiation era,
\begin{equation}
\label{eq:Tcut}
T_{\rm ini} < T_{\rm cut} = \sqrt{\frac{M_*}{2t_{\rm cut}}} \simeq 330\,\textrm{keV}\left(\frac{10^{-20}}{G\mu}\right)^{1/2} \,,
\end{equation}
where we used the standard time--temperature relation
\begin{equation}
\label{eq:HT2}
H = \frac{1}{2t} = \frac{T^2}{M_*} \,,
\end{equation}
during radiation domination. Here, $H$ denotes the Hubble rate and $M_* = [90/\left(\pi^2 g_*\right)]^{1/2}M_{\rm Pl}$ is a rescaled version of the reduced Planck mass that accounts for the effective number of relativistic degrees of freedom $g_*$. In Eq.~\eqref{eq:Tcut}, we set $g_* = 100$. In all analytical expressions below, we will continue to use this value, for simplicity, whereas we take into account the full temperature dependence of $g_*$ in all of our plots and numerical results. 

The temperature $T_{\rm cut}$ in Eq.~\eqref{eq:Tcut} is rather low. In fact, it is close to the temperature scale of big-bang nucleosynthesis, $T_{\rm BBN} \sim 1\,\textrm{MeV}$, and hence far away from the high energy scales that are typically associated with the production of cosmic strings. It is therefore not immediately obvious that the initial temperature $T_{\rm ini}$ can indeed be smaller than $T_{\rm cut}$, which is the necessary condition for the presence of the cutoff frequency $f_{\rm cut}$ in the GWB spectrum from fundamental loop oscillations. For this reason, we shall now discuss different possible choices for $T_{\rm ini}$ and argue that it is indeed possible to satisfy the condition $T_{\rm ini} < T_{\rm cut}$ across a large range of $G\mu$ values.

\subsection{Initial loop length}
\label{subsec:lini}

We define the initial time $t_{\rm ini}$ as the earliest time in the evolution of the string network when loop production is no longer impeded by thermal friction and GW emission is no longer subdominant to particle emission, i.e., $t_{\rm ini}$ marks the initial time from which on we rely on the standard computation of the GWB signal from cosmic strings. We do not expect significant GW production at earlier times.%
\footnote{See, however, Ref.~\cite{Mukovnikov:2024zed} for a recent analysis that discusses how, depending on the initial conditions of the network and its surroundings, GW emission during the friction regime can lead to a secondary peak in the GWB spectrum at high frequencies.}
This means in particular that $t_{\rm ini}$ determines the lower integration boundary in the integral over all possible times of GW emission from string loops that we must evaluate in order to compute the total GWB spectrum from cosmic strings; see Eq.~\eqref{eq:integral} below. In the literature, several possible choices for $t_{\rm ini}$ have been proposed; see Ref.~\cite{Gouttenoire:2019kij} for an extended discussion and Ref.~\cite{Servant:2023tua} for a recent review. In the following, we shall provide a brief and critical summary of these options.

\medskip\noindent\textbf{1) Network formation:} The string network forms during the cosmological phase transition that results in the spontaneous breaking of the underlying $U(1)$ symmetry. This phase transition takes place at temperatures around the symmetry breaking scale, $T_{\rm form} \sim v$, or equivalently, at times when $\rho_{\rm tot} = 3H^2 M_{\rm Pl}^2 \sim \mu^2$. In the following, we will work with the latter condition, for definiteness. During radiation domination, this condition yields
\begin{align}
T_{\rm form} & = \left(\frac{30}{\pi^2 g_*}\right)^{1/4}\mu^{1/2} \nonumber\\
& \simeq 5.1 \times 10^8\,\textrm{GeV}\left(\frac{G\mu}{10^{-20}}\right)^{1/2} \,,
\label{eq:Tform}
\end{align}
which corresponds to extremely early times,
\begin{equation}
t_{\rm form} = \frac{M_*}{2T_{\rm form}^2} \simeq 9.3 \times 10^{-25}\,\textrm{s}\left(\frac{10^{-20}}{G\mu}\right) \,,
\end{equation}
far away from the values of $t_{\rm cut}$ discussed in the previous subsection. The time scale $t_{\rm form}$ may in principle be regarded as the earliest possible choice for $t_{\rm ini}$; we, however, argue that this choice is unrealistic. In realistic microscopic models, we rather expect that the dynamics of the string network shortly after its formation are first dominated by thermal friction with the plasma through which the strings move and with which they interact. At early times, strings are thus unable to move freely, which means that the production of loops as well as GW emission from loops is initially strongly suppressed. 

\medskip\noindent\textbf{2) Thermal friction:} At a phenomenological level, friction can be accounted for in the equation of motion for Nambu--Goto strings in an expanding Universe in the form of a temperature-dependent friction term, $\beta\,T^3/\mu$, where $\beta$ is a model-dependent coefficient~\cite{Vilenkin_Shellard_2000, Everett:1981nj, deSousaGerbert:1988qzd, Alford_Wilczek_89}. In the following, we will work with $\beta \sim 1$ at temperatures above the electron--positron annihilation temperature, $T_e \sim 0.5\,\textrm{MeV}$, and $\beta = 0$ at $T \lesssim T_e$, when no more charge carriers are present in the thermal bath.%
\footnote{We thank J.\ J.\ Blanco-Pillado for a remark on this point.}
Comparing the thermal friction term to the Hubble friction term in the equation of motion, $2H$, allows us to estimate the temperature at the end of the friction regime, $T_{\rm fric}$, when the network begins to settle in the scaling regime that determines its evolution at all later times,
\begin{equation}
\label{eq:Tfric}
T_{\rm fric} = \frac{2\mu}{\beta M_*} \simeq 4.1\,\textrm{GeV} \left(\frac{1}{\beta}\right)\left(\frac{G\mu}{10^{-20}}\right) \,.
\end{equation}
As evident from Eqs.~\eqref{eq:Tform} and \eqref{eq:Tfric}, small values of $G\mu$ imply a large hierarchy $T_{\rm fric} \ll T_{\rm form}$, which results from the different power-law dependence on $G\mu$ of both temperatures, $T_{\rm form} \propto (G\mu)^{1/2}$ and $T_{\rm fric} \propto G\mu$. For small $G\mu$, the onset of the scaling regime is hence delayed, 
\begin{equation}
\label{eq:tfric}
t_{\rm fric} = \frac{M_*}{2T_{\rm fric}^2} \simeq 15\,\textrm{ns} \left(\frac{\beta}{1}\right)^2\left(\frac{10^{-20}}{G\mu}\right)^2 \,,
\end{equation}
which is much later than the time of network formation.

In some microscopic models, $t_{\rm fric}$ may serve as a reasonable choice for the initial time $t_{\rm ini}$. In Sec.~\ref{sec:vos}, we will therefore present two benchmark GWB spectra for parameter points that are based on the identification $t_{\rm ini} \rightarrow t_{\rm fric}$ and that, at the same time, manage to satisfy the condition $t_{\rm ini} > t_{\rm cut}$. In fact, from Eqs.~\eqref{eq:tcut} and \eqref{eq:tfric}, we can read off that the condition $t_{\rm fric} > t_{\rm cut}$ amounts to an upper bound on the string tension, $G\mu \lesssim 1.9 \times 10^{-23}$, below which a sharp cutoff frequency must appear in the $k=1$ contribution to the GWB spectrum. At such small $G\mu$ and setting $t_{\rm ini} \rightarrow t_{\rm fric}$, the loop length at $t_{\rm ini}$ is estimated as $\ell_{\rm fric} = \ell_*\left(t_{\rm fric}\right) = 2\alpha\,t_{\rm fric}$, which evaluates to
\begin{equation}
\ell_{\rm fric} = \frac{\alpha\beta^2M_*^3}{4\mu^2} \simeq 440\,\textrm{km} \left(\frac{\beta}{1}\right)^2\left(\frac{10^{-23}}{G\mu}\right)^2 \,,
\end{equation}
and which allows us to estimate the cutoff frequency $f_{\rm cut}$. For $G\mu$ values resulting in a clear hierarchy of time scales, $t_{\rm fric} \gg t_{\rm cut}$, a rough estimate can be obtained by neglecting the fact that loops shrink because of GW emission altogether, such that the exact expression in Eq.~\eqref{eq:fcut} turns into the simpler estimate $f_{\rm cut}^{\rm fric} \approx 2/\ell_{\rm fric}$, i.e.,
\begin{equation}
f_{\rm cut}^{\rm fric} \approx \frac{8\mu^2}{\alpha\beta^2M_*^3} \simeq 1.4\,\textrm{kHz} \left(\frac{1}{\beta}\right)^2\left(\frac{G\mu}{10^{-23}}\right)^2 \,.
\end{equation}
As we will see in Sec.~\ref{sec:vos}, where we also properly account for the temperature dependence of $g_*$, this estimate provides a reasonable approximation of the location of the cutoff frequency in the $k=1$ part of the GWB spectrum.

\medskip\noindent\textbf{3) Particle radiation from kink--kink collisions:} The end of the friction regime at $t_{\rm fric}$ only serves as a reasonable choice for the initial time $t_{\rm ini}$ in models in which string loops predominantly lose their energy via GW emission throughout their evolution. In many models, this is not the case and the energy loss of small loops at early times is actually dominated by particle emission~\cite{Blanco-Pillado:1998tyu,Olum:1998ag,Matsunami:2019fss,Auclair:2019jip,Blanco-Pillado:2023sap}. The reason for this is that small loops can feature a rich substructure, characterized by cusps and kinks that lift the topological stabilization of the string and thus give rise to bursts of particle radiation. 

Particle emission may, e.g., originate from the collisions of kinks propagating along strings that lead to a breakdown of the one-dimensional Nambu--Goto approximation and the production of particles with masses of the order of the symmetry breaking scale $v$. In Ref.~\cite{Matsunami:2019fss}, it was argued and numerically verified that the power radiated off in kink--kink collisions is roughly given by
\begin{equation}
\label{eq:Pkink}
P_{\rm kink} \sim \frac{N_k \epsilon}{\ell} \,,
\end{equation}
where $\epsilon \sim \mu^{1/2}$ is the $\ell$-independent energy emitted in a single radiation burst, and $N_k$ is the number of such bursts per oscillation period, which can assume values as large as $N_k \sim 10^3 \cdots 10^6$ in extreme cases~\cite{Binetruy:2010bq,Binetruy:2010cc,Ringeval:2017eww}.

The power in Eq.~\eqref{eq:Pkink} needs to be compared to the power of GW emission, which follows form multiplying Eq.~\eqref{eq:dldt} by the string tension, $P_{\rm GW} = \Gamma G\mu^2$. As evident from the expressions for $P_{\rm kink}$ and $P_{\rm GW}$, particle emission dominates over GW emission at early times, as long as all loops in the network are smaller than the critical length
\begin{equation}
\ell_{\rm kink} \sim \frac{N_k}{\Gamma G\mu^{3/2}} \simeq 320\,\textrm{nm} \left(\frac{N_k}{1}\right)\left(\frac{10^{-20}}{G\mu}\right)^{3/2} \,.
\end{equation}
Loops at least as large as $\ell_{\rm kink}$ form for the first time around $t_{\rm kink} = \ell_{\rm kink}/\left(2\alpha\right)$, which evaluates to
\begin{equation}
\label{eq:tkink}
t_{\rm kink} \sim \frac{N_k}{2\alpha\,\Gamma G\mu^{3/2}} \simeq 11\,\textrm{fs} \left(\frac{N_k}{1}\right)\left(\frac{10^{-20}}{G\mu}\right)^{3/2} \,,
\end{equation}
and which corresponds to temperatures of the order of
\begin{equation}
\label{eq:Tkink}
T_{\rm kink} = \sqrt{\frac{M_*}{2t_{\rm kink}}} \simeq 4.7\,\textrm{TeV} \left(\frac{1}{N_k}\right)^{1/2}\left(\frac{G\mu}{10^{-20}}\right)^{3/4} \hspace{-1em}.
\end{equation}

These estimates indicate that the time scale for particle emission from kink--kink collisions typically remains below the time scale of the friction regime, $t_{\rm kink} \ll t_{\rm fric}$, see Eqs.~\eqref{eq:tfric} and \eqref{eq:tkink}. This conclusion persists for a more extreme choice of $N_k$. Even for $N_k$ as large as $N_k \sim 10^6$, $t_{\rm kink}$ can become at most as long as the friction time scale, $t_{\rm kink} \lesssim t_{\rm fric}$, but not longer, for the $G\mu$ values we are interested in, $G\mu \sim 10^{-19}$. We therefore conclude that, for low-scale cosmic strings, particle emission from kink--kink collisions does not play any role during the scaling regime. For this reason, we shall discard the option to identify the initial time $t_{\rm ini}$ with $t_{\rm kink}$ in the following. 

\medskip\noindent\textbf{4) Particle radiation from cusps:} Besides kink--kink collisions, particle emission can also originate from cusps that form and propagate on string loops~\cite{Blanco-Pillado:1998tyu,Olum:1998ag}. Similarly as in the previous case, cusps represent regions where the Nambu--Goto approximation breaks down, because the topological stabilization of the string is lifted around the cusp, which paves the way for particle production. More precisely, cusps are regions where the string doubles back onto itself, which, together with its finite width, $r \sim 1/\sqrt{\mu} \sim 1/v$, results in the partial overlap of nearby parts of the string. Taking into account relativistic length contraction, the size of this overlap for a cusp on a loop of length $\ell$ is expected to be of the order of $\sqrt{rl}$~\cite{Blanco-Pillado:1998tyu,Olum:1998ag}. After a cusp has formed, the burst of particle radiation emanating from it causes the overlap region to shrink down to a size of the order of $r$. However, since we are interested in loops with $\ell \gg r$, this remaining size of the order of $r$ is practically negligible. This means that, effectively, particle emission from a cusp reduces the length of a loop from $\ell$ down to roughly $\ell - \sqrt{r\ell}$, which corresponds to a loss of energy of around $\mu \sqrt{r\ell} \sim \ell^{1/2}\mu^{3/4}$. Meanwhile, the authors of Ref.~\cite{Blanco-Pillado:2015ana} used a toy model to estimate that almost all loops develop $N_c \sim 1$ cusps per oscillation period $T\sim \ell$, implying an emitted power of
\begin{equation}
\label{eq:Pcusp}
P_{\rm cusp} \sim \frac{N_c\mu^{3/4}}{\ell^{1/2}} \,.
\end{equation}

Comparing $P_{\rm cusp}$ in Eq.~\eqref{eq:Pcusp} with the power of GW emission, $P_{\rm GW}$, shows that particle radiation dominates again over GW radiation at early times as long as all loops in the network are smaller than the critical length
\begin{equation}
\ell_{\rm cusp} \sim \frac{N_c^2}{\Gamma^2 G^2 \mu^{5/2}} \simeq 4.3\,\textrm{AU}\left(\frac{N_c}{1}\right)^2\left(\frac{10^{-20}}{G\mu}\right)^{5/2} \,.
\end{equation}
Loops at least as large as $\ell_{\rm cusp}$ form for the first time around $t_{\rm cusp} = \ell_{\rm cusp}/\left(2\alpha\right)$, which evaluates to
\begin{equation}
\label{eq:tcusp}
t_{\rm cusp} \sim \frac{N_c^2}{2\alpha\,\Gamma^2 G^2 \mu^{5/2}} \simeq 6.0\,\textrm{h} \left(\frac{N_c}{1}\right)^2\left(\frac{10^{-20}}{G\mu}\right)^{5/2} \hspace{-1em} ,
\end{equation}
and which corresponds to temperatures of the order of
\begin{equation}
\label{eq:Tcusp}
T_{\rm cusp} = \sqrt{\frac{M_*}{2t_{\rm cusp}}} \simeq 3.4\,\textrm{keV} \left(\frac{1}{N_c}\right)\left(\frac{G\mu}{10^{-20}}\right)^{5/4}  \,.
\end{equation}
Remarkably enough, this estimate falls below the temperature scales $T_{\rm cut}$ in Eq.~\eqref{eq:Tcut} and $T_{\rm fric}$ in Eq.~\eqref{eq:Tfric}. 
Therefore, if particle emission from cusps should indeed provide an efficient energy loss channel for string loops at early times, in line with the estimates in Refs.~\cite{Blanco-Pillado:1998tyu,Olum:1998ag,Blanco-Pillado:2015ana}, we are able to draw two conclusions: (i) $T_{\rm cusp} \ll T_{\rm fric}$ implies that we must not identify the initial time $t_{\rm ini}$ that enters the computation of the GWB spectrum with the end of the friction regime. While the network already begins to settle in the scaling regime around $t_{\rm fric}$, GW emission does not become the dominant energy loss mechanism until much later. In scenarios with dominant particle emission from cusps at early times, we should therefore identify $t_{\rm ini}$ with $t_{\rm cusp}$. (ii) $T_{\rm cusp} \ll T_{\rm cut}$ implies that string loops formed after $t_{\rm cusp}$ do not manage to reach zero length between the time of production and today. As a consequence, a sharp cutoff frequency appears in the GWB spectrum from fundamental loop oscillations, which we can estimate again by neglecting the decrease in loop length altogether, $f_{\rm cut}^{\rm cusp} \approx 2/\ell_{\rm cusp}$, i.e., 
\begin{equation}
\label{eq:fcutcusp}
f_{\rm cut}^{\rm cusp} \approx  \frac{2\,\Gamma^2 G^2 \mu^{5/2}}{N_c^2} \simeq 930\,\textrm{\textmu Hz} \left(\frac{1}{N_c}\right)^2\left(\frac{G\mu}{10^{-20}}\right)^{5/2} \,.
\end{equation}

The frequency estimate in Eq.~\eqref{eq:fcutcusp} falls into the mHz band, which indicates that the corresponding GWB signal may be observable with future space-borne GW experiments. In the next section, we will demonstrate that this is indeed the case, meaning that second-generation experiments like BBO and DECIGO will have a chance to measure the GWB spectrum from low-scale cosmic strings with a tension of around $G\mu \sim 10^{-19}$. 

Before we move on, we also point out that the hierarchy $t_{\rm cusp} \ll t_{\rm fric}$, which we encounter in the $t_{\rm ini} \rightarrow t_{\rm cusp}$ scenario, can be regarded as an advantage in the computation of the GWB spectrum that renders the numerical results of the standard formalism slightly more reliable than in the $t_{\rm ini} \rightarrow t_{\rm fric}$ scenario. To see this, note that the identification $t_{\rm ini} \rightarrow t_{\rm fric}$ always introduces some uncertainties related to the fact that, in this scenario, the computation of the GWB spectrum assumes the string network to be fully settled in the scaling regime for all times $t \geq t_{\rm fric}$, even though this assumption is not satisfied around the transition time $t_{\rm fric}$. The computation of the GWB spectrum in the scenario of cusp domination at early times does not suffer from this shortcoming. In this case, the string network has already settled in the scaling regime long before GW emission finally becomes the dominant mechanism of energy loss around $t_{\rm cusp}$. It is therefore unproblematic to assume a scaling network right from the initial time $t_{\rm ini}$ in this scenario.

Meanwhile, one may wonder to what extent the details of the transition from particle radiation to GW radiation around $t_{\rm cusp}$ may affect the GWB spectrum. However, in order to model this transition in a time-resolved fashion, one has to resort to a nonscaling description of the network along the lines of Ref.~\cite{Auclair:2019jip}, which is beyond the scope of this work. We will return to this question in the future~\cite{Schmitz:2024abc} (see also the discussion in Sec.~\ref{sec:conclusions}). In the present work, we shall, however, restrict ourselves to a simpler description that treats the time scales discussed above ($t_{\rm form}$, $t_{\rm fric}$, $t_{\rm kink}$, $t_{\rm cusp}$) as hard cutoffs~\cite{Gouttenoire:2019kij,Servant:2023tua}.

\begin{figure}
\begin{center}
\includegraphics[width = 0.48\textwidth]{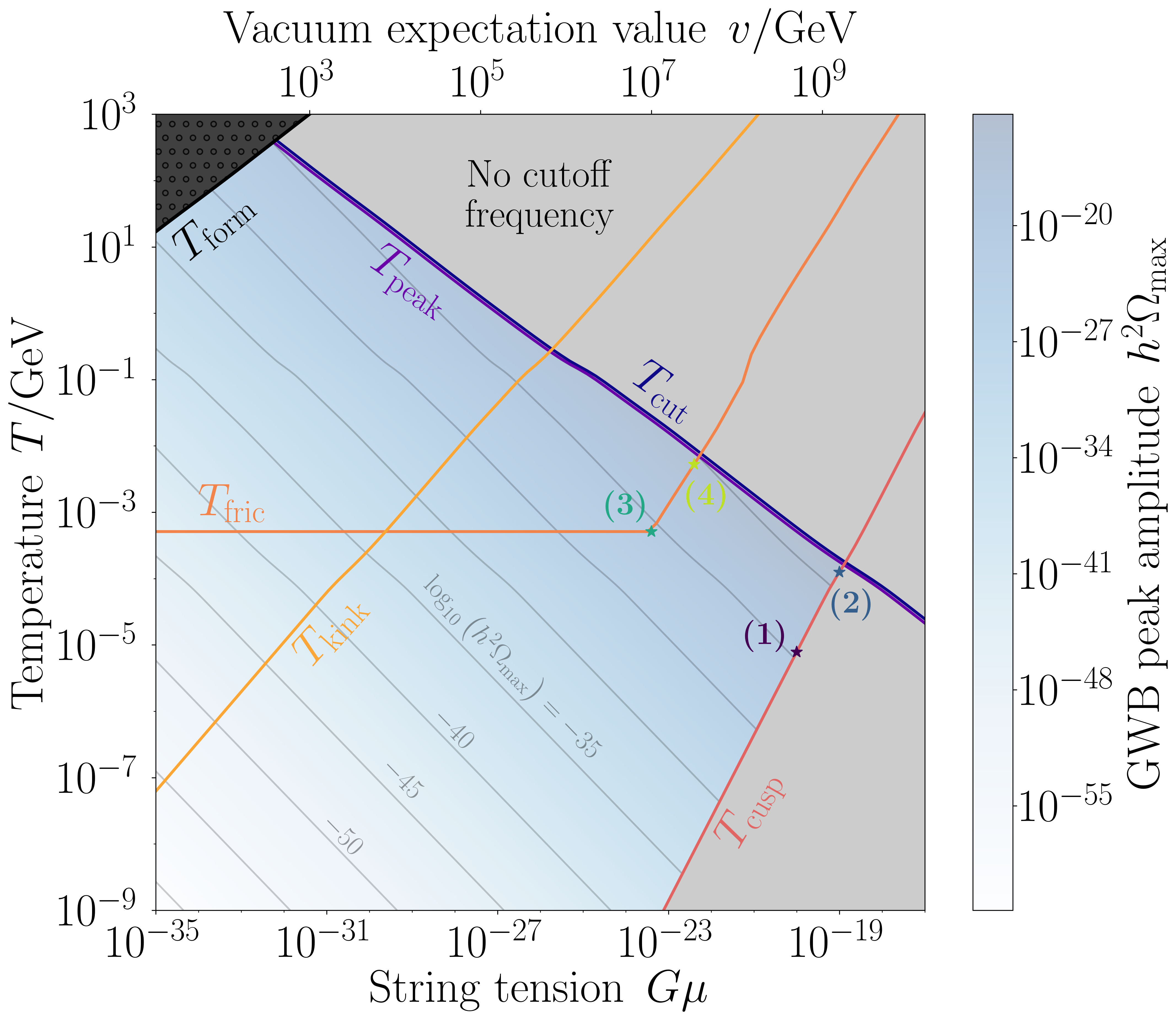}
\end{center}
\caption{Relevant temperature scales as functions of the string tension $G\mu$; see $T_{\rm cut}$ in Eq.~\eqref{eq:Tcut}, $T_{\rm form}$ in Eq.~\eqref{eq:Tform}, $T_{\rm fric}$ in Eq.~\eqref{eq:Tfric}, $T_{\rm kink}$ in Eq.~\eqref{eq:Tkink}, $T_{\rm cusp}$ in Eq.~\eqref{eq:Tcusp}, and $T_{\rm peak}$ in Eq.~\eqref{eq:Tpeak}. The blue shading and gray contours indicate the value of $h^2\Omega_{\rm max}$ in Eq.~\eqref{eq:Opeak} in dependence of $G\mu$ and $T_{\rm ini}$, where $T_{\rm ini}$ is identified with $T$ on the vertical axis. In realistic models, $T_{\rm ini}$ is not a free parameter, though, but must be identified with $T_{\rm form}$, $T_{\rm fric}$, $T_{\rm kink}$, or $T_{\rm cusp}$. As explained in the text, the identifications $T_{\rm ini} \rightarrow T_{\rm fric}$ and $T_{\rm ini} \rightarrow T_{\rm cusp}$ are particularly well motivated. In Sec.~\ref{sec:vos}, we discuss four benchmark points for these two scenarios, whose $G\mu$ and $T_{\rm ini}$ values are summarized in Table~\ref{tab:points} and shown above.}
\label{fig:Tscales}
\end{figure}

These remarks conclude our discussion of $f_{\rm cut}$ and $\ell_{\rm ini}$ in this section. A graphical summary of our results for the temperature scales $T_{\rm form}$, $T_{\rm fric}$, $T_{\rm kink}$, and $T_{\rm cusp}$ in the parameter region of interest is shown in Fig.~\ref{fig:Tscales}. Next, we are going to explicitly calculate the GWB spectrum from low-scale cosmic strings in the VOS model. 

%%%%%%%%%%%%%%%%%%%%%%%%%%%%%%%%%%%%%%%%%%%%%%%%%%%%%%%%%%%%%%%%%%%%%%%%%%%%%%%%%%%%%%%%%%%%%%%%%%%%

\section{GWB spectrum in the VOS model}
\label{sec:vos}

We express the GWB signal from cosmic strings in terms of the GW energy density power spectrum~\cite{Maggiore:1999vm,Caprini:2018mtu},
\begin{equation}
\label{eq:OGW}
\Omega_{\rm GW}\left(f\right) = \frac{1}{\rho_{\rm crit}}\frac{d \rho_{\rm GW}\left(f\right)}{d\,\ln f} \,.
\end{equation}
Here, $d\rho_{\rm GW}/d\ln f$ is the GW energy density per logarithmic frequency interval, and $\rho_{\rm crit} = 3H_0^2 M_{\rm Pl}^2$ denotes the critical energy density of the Universe, where $H_0 = 100\,h\,\textrm{km}/\textrm{s}/\textrm{Mpc}$ is the present-day value of the Hubble rate. The dimensionless Hubble rate is $h \simeq 0.7$; however, in our numerical results below, we will only work with $h^2\Omega_{\rm GW}$ instead of $\Omega_{\rm GW}$. Thus, all of our results will be independent of the exact value of $h$.

\subsection{Numerical results}

Based on the definition in Eq.~\eqref{eq:OGW}, the GWB spectrum emitted by the string network can be written as~\cite{Vachaspati:1984gt,Blanco-Pillado:2017oxo}
\begin{equation}
\label{eq:OGW2}
\Omega_{\rm GW}\left(f\right) = \frac{8\pi}{3H_0^2}\left(G\mu\right)^2\sum_{k=1}^{k_{\rm max}} P_k\,\mathcal{I}_k\left(f\right) \,,
\end{equation}
where the $k$ sum runs over all harmonic loop excitations up to some $k_{\rm max}$. We set $k_{\rm max} = 10^5$, in which case it is still numerically feasible to evaluate the sum. The physical cutoff in the $k$ sum is, however, located at much larger $k$ values, as we will discuss in more detail below.

The factor $P_k$ quantifies the GW power per $k$-mode,
\begin{equation}
\label{eq:Pk}
P_k = \frac{\Gamma}{H_{k_{\rm max}}^q} \frac{1}{k^q} \,,
\end{equation}
where $H_{k_{\rm max}}^q$ is the $k_{\rm max}$-th generalized harmonic number of order $q$, which ensures that the total emitted power, i.e., the sum over $P_k$, is normalized to $\Gamma$ [see Eq.~\eqref{eq:dldt}],
\begin{equation}
H_{k_{\rm max}}^q = \sum_{k=1}^{k_{\rm max}} \frac{1}{k^q} \,, \qquad \sum_{k=1}^{k_{\rm max}} P_k = \Gamma \,.
\end{equation}
The power $q$ in Eq.~\eqref{eq:Pk} reflects the dominant form of GW emission from string loops. In the following, we will set $q = \sfrac{4}{3}$, which corresponds to GW emission in the form of GW bursts from cusps on string loops~\cite{Vachaspati:1984gt, Damour:2001bk, Binetruy:2009vt}. As shown in Ref.~\cite{Blanco-Pillado:2017oxo}, this choice approximates the result for the power $P_k$ found in numerical simulations quite well, especially, at large $k$ values. By setting $q=\sfrac{4}{3}$, we also implicitly assume that particle emission from cusps can be neglected as long as $P_{\rm GW} \gg P_{\rm cusp}$ is satisfied, which means that newly formed loops lose their energy much faster via GW emission than particle emission. 

\begin{table}
\caption{Choice of the string tension $G\mu$ and initial temperature $T_{\rm ini}$ for the four benchmark points in Figs.~\ref{fig:Tscales} and \ref{fig:4spectra}.}
\label{tab:points}
\begin{center}
\renewcommand{\arraystretch}{1.5}
\begin{tabular}{|c c c|} 
\hline
Benchmark point & $\log_{10}\left(G\mu\right)$ & $\log_{10}\left(T_{\rm ini}/{\rm GeV}\right)$ \\ [0.5ex] 
\hline\hline
$1$ & $-20.0$ & $-5.107$ \\
\hline
$2$ & $-19.0$ & $-3.903$  \\
\hline
$3$ & $-23.4$ & $-3.292$ \\
\hline
$4$ & $-22.4$ & $-2.277$  \\
\hline
\end{tabular}
\end{center}
\end{table}

Finally, the frequency-dependent factor $\mathcal{I}_k$ in Eq.~\eqref{eq:OGW2} represents an integral of the loop number density $n\left(\ell,t\right)$,
\begin{equation}
\label{eq:integral}
\mathcal{I}_k\left(f\right) = \frac{2k}{f} \int_{t_{\rm ini}}^{t_0} \textrm{d}t \left(\frac{a\left(t\right)}{a_0}\right)^5 n\left(\frac{2k}{f}\frac{a\left(t\right)}{a_0},t\right) \,.
\end{equation}
This expression now shows what we already stated at the beginning of Sec.~\ref{subsec:lini}, namely, that the initial time $t_{\rm ini}$ marks the lower integration boundary in the integral over all possible times of GW emission. Similarly, the first argument of the loop number density $n\left(\ell,t\right)$ in the integrand, $\ell = 2k/f\left[a\left(t\right)/a_0\right]$, precisely reflects the relation in Eq.~\eqref{eq:flk}. The loop number density itself can be computed based on the VOS model~\citep{Sousa:2013aaa},
\begin{equation}
\label{eq:nloop}
n\left(\ell,t\right) = \mathcal{F}\:\frac{C_*\,\Theta\left(t-t_*\right)\Theta\left(t_*-t_{\rm ini}\right)}{y_*\left(y_* + \Gamma G\mu + \dot{y}_* t_*\right)t_*^4} \left(\frac{a\left(t_*\right)}{a\left(t\right)}\right)^3 \,,
\end{equation}
where we introduced the scaling variable $y = \ell/t$, which is closely related to the scaling variable $x=\ell/d_h$ introduced in Sec.~\ref{sec:intro}. $y_*$ specifically denotes the value of $y$ at the time of loop formation, $y_* = \ell_*/t_*$, where $t_*$ follows from solving the implicit relation
\begin{equation}
\label{eq:tstar}
t_* = \frac{\ell + \Gamma G\mu\,t}{y_* + \Gamma G \mu} \,, \qquad y_* = y\left(t_*\right) \,.
\end{equation}
The time dependence of $y$ can be expressed in terms of the correlation length $L$ of the long-string network,
\begin{equation}
y\left(t\right) = \alpha_L\,\xi\left(t\right) \,, \qquad L\left(t\right) = \xi\left(t\right)t \,,
\end{equation}
where $\alpha_L$ is a constant that we set to $\alpha_L = 0.37$, in agreement with numerical simulations~\citep{Martins:2000cs,Blanco-Pillado:2011egf,Blanco-Pillado:2013qja}. 

\begin{figure}
\begin{center}
\includegraphics[width = 0.48\textwidth]{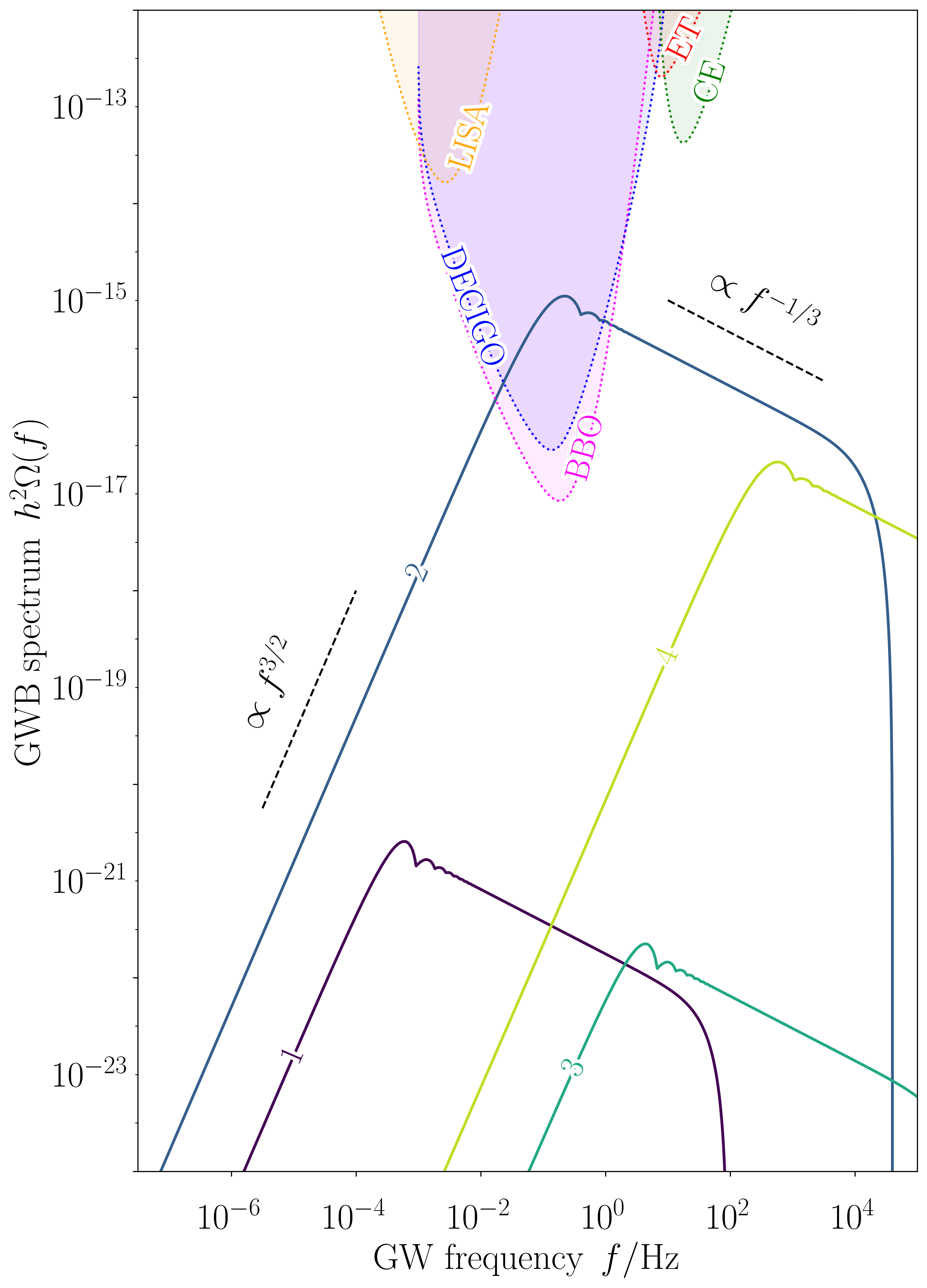}
\end{center}
\caption{Numerical GWB spectra based on the VOS model for the four benchmark points indicated in Fig.~\ref{fig:Tscales}; see Table~\ref{tab:points} for the respective values of $G\mu$ and $T_{\rm ini}$. Benchmark points 1 and 2 assume the identification $T_{\rm ini} \rightarrow T_{\rm cusp}$, while benchmark points 3 and 4 assume the identification $T_{\rm ini} \rightarrow T_{\rm fric}$. The filled curves at the top are power-law-integrated sensitivity curves for future experiments and are taken from Ref.~\cite{Schmitz:2020syl}.}
\label{fig:4spectra}
\end{figure}

The overall prefactors in Eq.~\eqref{eq:nloop}, $\mathcal{F}$ and $C_*$, are two numerical factors that account for the efficiency of GW production and the efficiency of loop production from the network, respectively. We follow Refs.~\citep{Blanco-Pillado:2013qja,Auclair:2019wcv} and set $\mathcal{F} = 0.1$. Meanwhile, the factor $C_*$ follows from
\begin{equation}
\label{eq:Cstar}
C_* = C\left(t_*\right) \,, \qquad C\left(t\right) = \frac{\tilde{c}}{\sqrt{2}}\frac{\bar{v}\left(t\right)}{\xi^3\left(t\right)} \,,
\end{equation}
where $\tilde{c} \simeq 0.23$ is the so-called loop-chopping parameter; similar to $\alpha_L$, it needs to be extracted from numerical simulations~\citep{Martins:2000cs,Blanco-Pillado:2011egf,Blanco-Pillado:2013qja}. $\bar{v}$ denotes the root-mean-square velocity of the long strings. Its time dependence, alongside the time dependence of the dimensionless correlation length $\xi$, is described by the VOS equations. We numerically solve these equations, which provides us with the explicit time dependence of $\bar{v}$ and $\xi$ and hence the explicit time dependence of $C$ and $y$. In this numerical analysis, we explicitly take into account the exact evolution of the scale factor $a$ and the full temperature dependence of $g_*$.

The machinery in Eqs.~\eqref{eq:OGW2} to \eqref{eq:Cstar} allows us to compute the GWB spectrum from cosmic strings for any choice of the string tension $G\mu$ and initial temperature $T_{\rm ini}$. As explained in the previous section, we are mostly interested in scenarios based on the identification $T_{\rm ini} \rightarrow T_{\rm fric}$ or $T_{\rm ini} \rightarrow T_{\rm cusp}$. For each of these two cases, we consider two explicit benchmark points whose location in parameter space is indicated in Fig.~\ref{fig:Tscales} and whose respective values of $G\mu$ and $T_{\rm ini}$ are listed in Table~\ref{tab:points}. At this point, note that, once $T_{\rm ini}$ has been identified with a particular temperature scale, it is no longer an independent parameter but rather a function of $G\mu$. This means that the $G\mu$ values in Table~\ref{tab:points} are freely chosen; the $T_{\rm ini}$ values in the last column, however, follow from evaluating $T_{\rm fric}$ or $T_{\rm cusp}$ at the respective values of $G\mu$. The four benchmark points shown in Fig.~\ref{fig:Tscales} roughly span the phenomenologically interesting parameter region. Parameter points in between these benchmark points might, e.g., be realized in scenarios where particle emission from string cusps is less efficient than what is described by Eq.~\eqref{eq:Pcusp}~\cite{Olum:1998ag}, such that $T_{\rm cusp}$ moves closer to $T_{\rm fric}$.

The GWB spectra for our four benchmark points are shown in Fig.~\ref{fig:4spectra}. As discussed in Sec.~\ref{sec:intro}, they display indeed a characteristic sequence of peaks and dips close to their respective maxima. We emphasize that the four GWB spectra in Fig.~\ref{fig:4spectra} follow straightforwardly from the standard VOS framework for the GW signal from cosmic strings and our choice of parameter values in Table~\ref{tab:points}. No extra ingredient or assumption beyond what we already discussed is required in order to arrive at these signal predictions for low-scale strings. This observation is the main result of the present paper. The VOS model predicts a GW signal from low-scale strings that is qualitatively very different from the GW signal from high-scale strings. For a favorable choice of parameter values, the GW signal from low-scale strings, including the sequence of peaks and dips in the spectrum, might be even observable by future experiments such as BBO and DECIGO.

\subsection{Analytical results}

The main properties of the GWB spectra in Fig.~\ref{fig:4spectra} can be understood analytically. To show that this is indeed the case, we first return to the expression for the GWB spectrum in Eq.~\eqref{eq:OGW2} and write the sum over $k$-modes as
\begin{equation}
\label{eq:OGW3}
\Omega_{\rm GW}\left(f\right) = \frac{1}{H_{k_{\rm max}}^q} \sum_{k=1}^{k_{\rm max}} \frac{\Omega_{\rm GW}^{(1)}\left(f/k\right)}{k^q} \,,
\end{equation}
where $\Omega_{\rm GW}^{(1)}$ is the GWB spectrum from the $k=1$ mode,
\begin{equation}
\label{eq:Omega1}
\Omega_{\rm GW}^{(1)}\left(f\right) = \frac{8\pi}{3H_0^2}\left(G\mu\right)^2 \Gamma\,\mathcal{I}_1\left(f\right) \,.
\end{equation}
The relation in Eq.~\eqref{eq:OGW3}, which immediately follows from the $k$ dependence of $P_k$ in Eq.~\eqref{eq:Pk} and the $k$ dependence of $\mathcal{I}_k$ in Eq.~\eqref{eq:integral}, represents an important property of the GWB spectrum from cosmic strings, namely, that the contributions from higher $k$-modes can be readily expressed in terms of the $k=1$ contribution, evaluated at the rescaled frequency $f/k$ and multiplied by a simple $k$-dependent factor. This property is of central importance for the GWB spectrum from low-scale strings.

Next, let us evaluate $\Omega_{\rm GW}^{(1)}$ analytically, which, in principle, receives three different contributions: GW emission during the radiation era, from loops that form during the radiation era (RR); GW emission during the matter era, from loops that form during the radiation era (RM); and GW emission during the matter era, from loops that form during the matter era (MM). Out of these contributions to the spectra in Fig.~\ref{fig:4spectra}, the only relevant one is the RM contribution. The RR contribution is suppressed in the case of low-scale strings, which reflects the fact that, for $\Gamma G\mu \lll 1$, there is not enough time for loops formed during the radiation era to lose any appreciable amount of energy because of GW emission during the radiation era; see Eq.~\eqref{eq:dldt}. Meanwhile, the MM contribution is suppressed by the small number density of MM loops. Using our numerical code for the GWB spectrum, we are able to corroborate this understanding and confirm that it is indeed sufficient to only keep the RM contribution.

In order to evaluate the RM contribution to $\Omega_{\rm GW}^{(1)}$, we require the number density of loops that form during the radiation era, evaluated during the matter era, $n_{\rm RM}$, which can be derived from Eq.~\eqref{eq:nloop}. To do so, we assume pure radiation domination right until matter--radiation equality, i.e., at all times $t < t_{\rm eq} \simeq 50,000\,\textrm{yr}$, followed by pure matter domination starting immediately at $t_{\rm eq}$. Moreover, we neglect again any changes in the relativistic number of degrees of freedom, $g_*$. We thus obtain
\begin{equation}
\label{eq:nRM}
n_{\rm RM}\left(\ell,t\right) = \frac{\mathcal{F}\,C_*\,\Theta}{y_*\left(y_* + \Gamma G\mu\right)t_*^4} \left(\frac{t_*}{t_{\rm eq}}\right)^{3/2}\left(\frac{a_{\rm eq}}{a\left(t\right)}\right)^3 .
\end{equation}
Here, $\Theta$ is a shorthand notation for three Heaviside functions, $\Theta\left(t-t_{\rm eq}\right)\Theta\left(t_{\rm eq} - t_*\right)\Theta\left(t_*-t_{\rm ini}\right)$, and $a_{\rm eq}$ is the scale factor at matter--radiation equality, $a_{\rm eq} \simeq a_0/3400$, where $a_0$ is the present-day value of the scale factor. At high temperatures, deep in the radiation-dominated era, the VOS equations yield $\xi_r \simeq 0.27$ and $\bar{v}_r \simeq 0.66$, which implies that $y_*$ assumes a fixed value, $y_* = y_r \simeq 0.10$. For constant $y_* = y_r$, Eq.~\eqref{eq:tstar} then turns from an implicit relation to an explicit expression for $t_*$, which we can use to simplify the RM loop number density in Eq.~\eqref{eq:nRM}, 
\begin{equation}
\label{eq:nRMlt}
n_{\rm RM}\left(\ell,t\right) = \frac{\mathcal{F}\,C_r\,\Theta\left(y_r + \Gamma G \mu\right)^{3/2}}{t_{\rm eq}^{3/2}y_r\left(\ell + \Gamma G\mu\,t\right)^{5/2}}\left(\frac{a_{\rm eq}}{a\left(t\right)}\right)^3 \,,
\end{equation}
where $C_r = \tilde{c}\,\bar{v}_r/\left(\sqrt{2}\,\xi_r^3\right) \simeq 5.5$. Finally, for $\Gamma G\mu \ll y_r$ (the only case we are interested in), $n_{\rm RM}$ simplifies to
\begin{equation}
\label{eq:nRM2}
n_{\rm RM}\left(\ell,t\right) \simeq \frac{A_r\,\Theta}{t_{\rm eq}^{3/2}\left(\ell + \Gamma G\mu\,t\right)^{5/2}}\left(\frac{a_{\rm eq}}{a\left(t\right)}\right)^3 \,,
\end{equation}
whose normalization, $A_r = \mathcal{F}\,C_r\,y_r^{1/2}\simeq 0.17$, agrees with the normalization of the number density in Eq.~(3.20) in Ref.~\cite{Auclair:2019wcv} (see also Ref.~\cite{Blanco-Pillado:2024aca} for a recent update).%
\footnote{Eq.~(3.20) in Ref.~\cite{Auclair:2019wcv} does not contain any Heaviside functions; the hierarchy of time scales, $t_{\rm ini} < t_* < t_{\rm eq} < t$, therefore needs to be imposed by hand in this equation. Its mass dimension is, moreover, incorrect. This, however, appears to be a mere typo and can be remedied by adding a factor of $H_0^{3/2}$ in the numerator.}

With the result for $n_{\rm RM}$ in Eq.~\eqref{eq:nRM2} at hand and working with the scale factor in pure matter domination, 
\begin{equation}
\label{eq:atm}
a\left(t\right) = a_0\left(\frac{3}{2}\,\Omega_{\rm m}^{1/2}H_0\,t\right)^{2/3} \,, \qquad h^2\Omega_{\rm m} \simeq 0.14 \,,
\end{equation}
we are now able to evaluate the integral in Eq.~\eqref{eq:integral} analytically and compute the resulting GWB spectrum. At low frequencies, we find a $f^{3/2}$ power-law behavior,
\begin{align}
\label{eq:Omegalow}
& h^2\Omega_{\rm low}^{(1)}\left(f\right) \approx \mathcal{A}_{\rm low} f^{3/2} \nonumber\\
& \simeq 7.7 \times 10^{-23}\left(\frac{G\mu}{10^{-20}}\right)^2\left(\frac{f}{10^{-5}\,\textrm{Hz}}\right)^{3/2} \,.
\end{align}
Here, we introduced $\mathcal{A}_{\rm low}$ for later convenience,
\begin{equation}
\mathcal{A}_{\rm low} = \frac{\mathcal{A}}{2^{5/2}\,H_{\rm m}} \left(1-\frac{a_{\rm eq}}{a_0}\right) \,,
\end{equation}
where $H_{\rm m} = \Omega_{\rm m}^{1/2}H_0$ and $\mathcal{A}$ is a shorthand notation for
\begin{equation}
\mathcal{A} = \frac{16\pi\left(G\mu\right)^2 \Gamma}{3\left(H_0/h\right)^2}\,\frac{A_r\left(a_{\rm eq}/a_0\right)^3}{t_{\rm eq}^{3/2}} \,.
\end{equation}
Next, we compute the high-frequency part of the spectrum.  In doing so, let us assume for a moment a sufficiently large $G\mu$ value, such that no cutoff appears in the spectrum. Expanding the GWB spectrum in Eq.~\eqref{eq:Omega1} at high frequencies then yields a $f^{-1}$ power-law behavior,
\begin{align}
\label{eq:Omegahigh}
& h^2\Omega_{\rm high}^{(1)}\left(f\right) \approx \mathcal{A}_{\rm high} f^{-1} \nonumber\\
& \simeq 9.5 \times 10^{-17}\left(\frac{10^{-20}}{G\mu}\right)^{1/2}\left(\frac{10^5\,\textrm{Hz}}{f}\right)^{-1} \,,
\end{align}
where $\mathcal{A}_{\rm high}$ is again proportional to the parameter $\mathcal{A}$,
\begin{equation}
\mathcal{A}_{\rm high} = 4\,\mathcal{A}\,H_{\rm m}^{3/2}\left(\frac{3}{2\,\Gamma G\mu}\right)^{5/2}\left[\left(\frac{a_0}{a_{\rm eq}}\right)^{1/4}-1\right] \,.
\end{equation}

At intermediate frequencies, the GWB spectrum exhibits a broken power-law peak, where the asymptotic $f^{3/2}$ behavior at low frequencies smoothly transitions to the asymptotic $f^{-1}$ behavior at high frequencies. We are able to estimate the position of the peak frequency $f_{\rm peak}$ by equating the expressions in Eqs.~\eqref{eq:Omegalow} and \eqref{eq:Omegahigh},
\begin{equation}
h^2\Omega_{\rm low}^{(1)}\left(f_{\rm peak}\right) = h^2\Omega_{\rm high}^{(1)}\left(f_{\rm peak}\right) \,.
\end{equation}
Solving this relation for $f_{\rm peak}$ results in
\begin{equation}
\label{eq:fpeak}
f_{\rm peak} = \frac{3H_{\rm m}}{\Gamma G\mu} \left[4\,\frac{\left(a_0/a_{\rm eq}\right)^{1/4}-1}{1-\left(a_{\rm eq}/a_0\right)}\right]^{2/5} \,,
\end{equation}
which evaluates to $f_{\rm peak} \simeq 27\,\textrm{Hz}$ for $G\mu = 10^{-20}$.

The peak frequency in Eq.~\eqref{eq:fpeak} needs to be compared to the cutoff frequency $f_{\rm cut}$. If $f_{\rm cut} < f_{\rm peak}$, the broken power-law peak is actually absent, and the GWB spectrum is well described by a single $f^{3/2}$ power law that abruptly drops to zero at $f_{\rm cut}$. The condition $f_{\rm cut} < f_{\rm peak}$ can be rewritten as a lower limit on the initial time $t_{\rm ini}$,
\begin{equation}
\label{eq:tpeak}
t_{\rm ini} > t_{\rm peak} = \frac{1}{\alpha f_{\rm peak}} + t_{\rm cut} \,,
\end{equation}
where we made again use of the fact that $t_0 \ggg t_{\rm ini}$. Thanks to Eq.~\eqref{eq:HT2}, this lower bound on $t_{\rm ini}$ can be formulated as an upper bound on the temperature scale $T_{\rm ini}$,
\begin{equation}
\label{eq:Tpeak}
T_{\rm ini} < T_{\rm peak} = \sqrt{\frac{M_*}{2t_{\rm peak}}} \,.
\end{equation}
In Fig.~\ref{fig:Tscales}, we plot $T_{\rm peak}$ as a function of $G\mu$, which illustrates that $T_{\rm peak}$ is always slightly smaller than $T_{\rm cut}$. Of course, this immediately follows from the fact that the condition $t_{\rm ini} > t_{\rm peak}$ in Eq.~\eqref{eq:tpeak} is always slightly stronger than the condition $t_{\rm ini} > t_{\rm cut}$ in Eq.~\eqref{eq:tcut}. In the following, we shall focus on the stronger of these two conditions, $t_{\rm ini} > t_{\rm peak}$, as it results in the more pronounced cutoff in the GWB spectrum. That is, if $t_{\rm ini} > t_{\rm peak}$, 
\begin{equation}
\label{eq:drop}
h^2\Omega^{(1)}\left(f\right) \approx \Theta\left(f_{\rm cut}-f\right)\,h^2\Omega_{\rm low}^{(1)}\left(f\right) \,.
\end{equation}
Note that the four benchmark points in Fig.~\ref{fig:Tscales} do satisfy the stronger condition $T_{\rm ini} < T_{\rm peak}$. Because of the close proximity of the scales $T_{\rm cut}$ and $T_{\rm peak}$, parameter points in the thin strip of parameter space where $T_{\rm cut} > T_{\rm ini} > T_{\rm peak}$ would, on the other hand, require a fine-tuned value of $G\mu$; we will not consider this possibility further.

The expression in Eq.~\eqref{eq:drop} marks our final result for the GWB spectrum from the fundamental oscillation mode. Inserting this result into Eq.~\eqref{eq:OGW3} allows us to compute the total GWB spectrum originating from all modes,
\begin{equation}
h^2\Omega_{\rm GW}\left(f\right) \approx \frac{\mathcal{A}_{\rm low}}{H_{k_{\rm max}}^q} \sum_{k = k_{\rm min}}^{k_{\rm max}} \frac{1}{k^q}\left(\frac{f}{k}\right)^{3/2} \,,
\end{equation}
where the sum now starts at $k_{\rm min} = \lceil f/f_{\rm cut}\rceil$, i.e., the first integer larger than $f/f_{\rm cut}$. The discrete sum over all $k$-modes can be written in terms of the Hurwitz zeta function $\zeta\left(p,a\right) = \sum_{k=0}^{\infty}1/\left(k+a\right)^p$, which yields
\begin{equation}
h^2\Omega_{\rm GW}\left(f\right) \approx \frac{\mathcal{A}_{\rm low}}{H_{k_{\rm max}}^q}\,f^{3/2}\, \Sigma \,,
\end{equation}
where we introduced $\Sigma$ as a shorthand notation for
\begin{equation}
\Sigma = \zeta\left(q+\sfrac{3}{2},k_{\rm min}\right) - \zeta\left(q+\sfrac{3}{2},k_{\rm max}+1\right) \,.
\end{equation}
At $f \leq f_{\rm cut}$, the function $\Sigma$ assumes a constant value,
\begin{equation}
k_{\rm min} =1 : \quad \Sigma = H_{k_{\rm max}}^{q+\sfrac{3}{2}} \,,
\end{equation}
which means that, at low frequencies, the total GWB spectrum coincides with the spectrum from the fundamental oscillation mode, up to a global prefactor,
\begin{equation}
h^2\Omega_{\rm low}\left(f\right) \approx \frac{H_{k_{\rm max}}^{q+\sfrac{3}{2}}}{H_{k_{\rm max}}^q}\,\mathcal{A}_{\rm low}\,f^{3/2} \,.
\end{equation}

In particular, for $q=\sfrac{4}{3}$ and $k_{\rm max} = 10^5$, we find
\begin{equation}
h^2\Omega_{\rm low}\left(f\right) \approx 0.35\,\mathcal{A}_{\rm low}\,f^{3/2} \,,
\end{equation}
which confirms the $f^{3/2}$ power-law behavior at low frequencies that we observe in Fig.~\ref{fig:4spectra}.
Evaluating this expression at the cutoff frequency $f_{\rm cut}$ provides us with an estimate of the peak amplitude in the GWB spectrum,
\begin{equation}
\label{eq:Opeak}
h^2\Omega_{\rm max} \approx 0.35\,\mathcal{A}_{\rm low}\,f_{\rm cut}^{3/2} \,.
\end{equation}
This estimate is a function of two parameters: the string tension $G\mu$ and, via the dependence of $f_{\rm cut}$ on $t_{\rm ini}$ [see Eq.~\eqref{eq:fcut}], the initial temperature $T_{\rm ini}$. In any given realistic scenario, $T_{\rm ini}$ needs to be identified with one of the temperature scales discussed in Sec.~\ref{subsec:lini}, rendering it a function of $G\mu$ after all. Still, it is instructive to forget about this point for a moment and simply treat $G\mu$ and $T_{\rm ini}$ as two independent parameters that determine the peak amplitude $h^2\Omega_{\rm max}$. In Fig.~\ref{fig:Tscales}, we do exactly that and show contour lines of $h^2\Omega_{\rm max}$ in dependence of the $G\mu$ values on the horizontal axis and the $T_{\rm ini}$ values on the vertical axis. From these contour lines, we conclude that the four benchmark points listed in Table~\ref{tab:points} are representative of the relevant region in parameter space. At smaller values of $G\mu$ or $T_{\rm ini}$, the GWB signal will be even weaker, far away from the sensitivity reach of future GW experiments. The contour plot of $h^2\Omega_{\rm max}$ shows in particular that benchmark point 2 is indeed located in the sweet spot of parameter space where the GWB signal from low-scale strings may be within the reach of DECIGO and BBO. Small variations in $G\mu$ or $T_{\rm ini}$ around this point will equally lead to observable GWB signals. In that sense, benchmark point 2 is not fine-tuned. At the same time, it is clear that the most interesting region of parameter space is rather small\,---\,GWs from low-scale strings will only be observable in future measurements in a range of optimistic scenarios.

\begin{figure}
\begin{center}
\includegraphics[width = 0.48\textwidth]{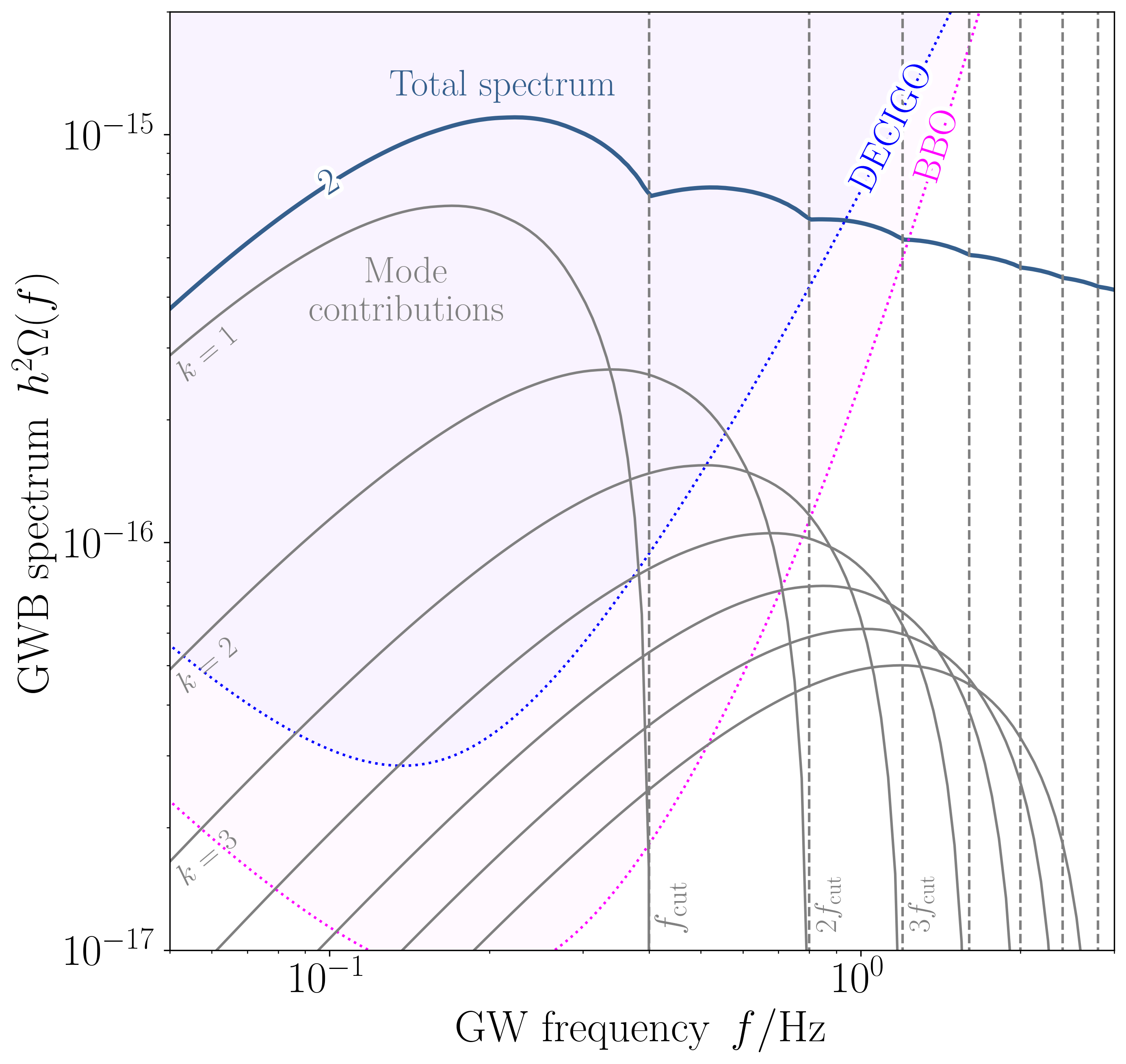}
\end{center}
\caption{GWB spectrum for benchmark point 2; see Table~\ref{tab:points} and Fig.~\ref{fig:4spectra}. The solid blue line shows the full spectrum, while the thin gray lines indicate the respective contributions from harmonic string oscillations with $k=1,2,3,\cdots$. The dashed vertical lines represent integer multiples of $f_{\rm cut}$ in Eq.~\eqref{eq:fcut}.}
\label{fig:benchmark}
\end{figure}

At frequencies resulting in the hierarchy $1 \ll k_{\rm min} \ll k_{\rm max}$, we can expand the $\Sigma$ factor in inverse powers of $f$,
\begin{equation}
1 \ll k_{\rm min} \ll k_{\rm max} : \quad \Sigma \approx \frac{1}{q+\sfrac{1}{2}} \left(\frac{f_{\rm cut}}{f}\right)^{q+\sfrac{1}{2}} \,,
\end{equation}
up to higher-order terms in $f_{\rm cut}/f$. At frequencies beyond the cutoff frequency $f_{\rm cut}$ in the GWB spectrum from the $k=1$ mode, the total GWB spectrum thus displays a power-law behavior with spectral index $1-q$,
\begin{equation}
h^2\Omega_{\rm tail}\left(f\right) \approx \frac{\mathcal{A}_{\rm low}\,f_{\rm cut}^{3/2}}{H_{k_{\rm max}}^q\left(q+\sfrac{1}{2}\right)}\left(\frac{f_{\rm cut}}{f}\right)^{q-1}  \,,
\end{equation}
which, for $q=\sfrac{4}{3}$ and $k_{\rm max} = 10^5$, evaluates to
\begin{equation}
h^2\Omega_{\rm tail}\left(f\right) \approx 0.15\,\mathcal{A}_{\rm low}\,f_{\rm cut}^{3/2}\left(\frac{f_{\rm cut}}{f}\right)^{1/3} \,,
\end{equation}
in excellent agreement with the $f^{-1/3}$ power-law behavior at large frequencies that we observe in Fig.~\ref{fig:4spectra}.

Finally, at frequencies resulting in $k_{\rm min} > k_{\rm max}$, the GWB spectrum vanishes. As a consequence, the $f^{-1/3}$ power-law behavior at frequencies $f_{\rm cut} \ll f \ll k_{\rm max}\,f_{\rm cut}$ suddenly transitions to a rapid drop-off as $f$ approaches $k_{\rm max}\,f_{\rm cut}$, which agrees again with the behavior that we observe in Fig.~\ref{fig:4spectra}. We thus conclude that the $f^{-1/3}$ power-law behavior at frequencies beyond $f_{\rm cut}$ spans across roughly $\log_{10}k_{\rm max}$ orders of magnitude in frequency space. In Fig.~\ref{fig:4spectra}, we set $k_{\rm max} = 10^5$, for which the numerical evaluation of the sum in Eq.~\eqref{eq:OGW3} is still computationally feasible. In realistic scenarios, the actual value of $k_{\rm max}$ may, however, be much larger. To see this, note that a string loop of length $\ell$ can support all harmonic oscillations with wavelengths $\lambda = \ell/\left(2k\right)$ that are sufficiently larger than the string diameter, $\lambda \gg 1/v$. This requirement translates into an upper limit on $k$,
\begin{equation}
k \ll \frac{\ell v}{2} \sim 10^{35}\left(\frac{\ell}{0.1\,\textrm{AU}}\right)\bigg(\frac{v}{10^9\,\textrm{GeV}}\bigg) \,,
\end{equation}
which is vastly larger than the value of $k_{\rm max} = 10^5$ chosen in our analysis. It is therefore conceivable that the $f^{-1/3}$ power-law behavior may actually continue to far higher frequencies than what is shown in Fig.~\ref{fig:4spectra}. For the parameter regions that we are concerned with in this paper, the high-frequency tail of the GWB spectrum is, however, not within the reach of near-future GW experiments, anyway.

Meanwhile, in Fig.~\ref{fig:benchmark}, we zoom in on the parts of the GWB spectrum belonging to benchmark point 2 that \textit{are} within the reach of near-future GW experiments, namely, BBO and DECIGO. This close-up serves as an illustrative summary of the main features that we predict to be present in the GWB spectrum from low-scale strings: At low frequencies, the spectrum displays an $f^{3/2}$ power-law behavior, before it transitions to an $f^{-1/3}$ power-law behavior around frequencies of the order of $f_{\rm cut}$. The GWB spectrum thus has the shape of a broken power law that reaches a peak amplitude of around $h^2\Omega_{\rm max}$ [see Eq.~\eqref{eq:Opeak}] at a frequency of around $f_{\rm cut}$ [see Eq.~\eqref{eq:fcut}]. On top, the discrete nature of the harmonic loop oscillations results in a modulation of this broken power law, characterized by a series of peaks and dips at frequencies $f \gtrsim f_{\rm cut}$. In particular, since the individual contributions to the total spectrum from loop oscillations with mode numbers $k=1,2,3,\cdots$ drop to zero at frequencies $kf_{\rm cut}$, respectively, the dips are precisely located at integer multiples of $f_{\rm cut}$. This integer sequence of dips on top of a broken power law with power-law indices of $\sfrac{3}{2}$ and $-\sfrac{1}{3}$ provides an exciting and well-defined target for future GW experiments in the $\mathcal{O}\left(0.1\cdots1\right)\,\textrm{Hz}$ frequency band.

\subsection{Loop number and energy densities}

The key assumption underlying our analysis in this paper is stated in Eq.~\eqref{eq:lt0}: we focus on string tensions that are so low that no string loop produced in the early Universe has had enough time yet to shrink to zero size because of GW emission. This means that all loops produced throughout the evolution of the string network are still around today, which raises the question whether they could cause any cosmological problems. To address this question, we shall now compute the total number and energy densities of string loops in the present Universe, which will allow us to conclude that the present-day abundance of string loops is cosmologically harmless.

The total loop number density simply follows from integrating $n\left(\ell,t\right)dl$, i.e., the number of loops per spatial volume whose length falls into the interval $\left[\ell,\ell+d\ell\right]$,
\begin{equation}
N\left(t\right) = \int_0^\infty d\ell\:n\left(\ell,t\right) \,.
\end{equation}
We are interested in $N\left(t\right)$ at times after matter--radiation equality, which means that the integrand in this expression receives two contributions: (i) $n_{\rm RM}$, the number density of loops born before $t_{\rm eq}$, evaluated at $t > t_{\rm eq}$, and (ii) $n_{\rm MM}$, the number density of loops born after $t_{\rm eq}$, evaluated at $t > t_{\rm eq}$. As discussed above, $n_{\rm MM}$ is, however, strongly suppressed. We are thus able to neglect $n_{\rm MM}$ in the computation of $N$ and simply work with $n_{\rm RM}$,
\begin{equation}
N\left(t\right) \approx \int_0^\infty d\ell\:n_{\rm RM}\left(\ell,t\right) \,.
\end{equation}
Using the explicit expression for $n_{\rm RM}$ in the VOS model in Eq.~\eqref{eq:nRMlt}, we are able to derive the following estimate,
\begin{equation}
N\left(t\right) \approx \frac{2/3}{y_r^{3/2}t_{\rm ini}^{3/2}}\frac{A_r}{t_{\rm eq}^{3/2}}\left(\frac{a_{\rm eq}}{a\left(t\right)}\right)^3 \,,
\end{equation}
up to numerically subleading corrections; for more details, see Ref.~\cite{Schmitz:2024abc2}, which discusses similar integrals over the loop number density in much greater detail. The current total loop number density is then roughly
\begin{equation}
N\left(t_0\right) \simeq \frac{49}{\textrm{kpc}^3} \left(\frac{10^2\,\textrm{s}}{t_{\rm ini}}\right)^{3/2} \,.
\end{equation}

Only a fraction of these loops is, however, of astrophysical size. To illustrate this, let us compute the current density of loops with a length of at least, say, $1\,\textrm{AU}$,
\begin{align}
N_{> 1\,\textrm{AU}}\left(t_0\right) & \approx \int_{1\,\textrm{AU}}^\infty d\ell\:n_{\rm RM}\left(\ell,t_0\right) \\
& \approx \frac{A_r}{t_{\rm eq}^{3/2}}\left(\frac{a_{\rm eq}}{a_0}\right)^3 \frac{2/3}{\left(1\,\textrm{AU} + \Gamma G\mu\, t_0\right)^{3/2}} \,,
\end{align}
which, for all $G\mu$ values of interest, $G\mu \lesssim 10^{-19}$, evaluates to roughly $0.14$ string loops per $\textrm{kpc}^3$. This number density appears to be unproblematic. At the same time, one may wonder if string loops in the direct vicinity of the solar system might lead to interesting phenomenological signatures, such as microlensing events~\cite{Vilenkin_1984,Mack:2007ae,Kuijken:2007ma,Bloomfield:2013jka} or GW bursts~\cite{Damour:2000wa,Damour:2001bk,Damour:2004kw,Siemens:2006vk}. We expect any such effect to be suppressed by the low string tensions that we are dealing with in this paper. Nonetheless, it might be interesting to revisit possible signatures of nearby low-scale strings in future work.

From a cosmological perspective, the energy density of string loops is more relevant than their number density. Neglecting again the MM contribution to $n$, we can write for the energy density parameter $\Omega$, i.e., the loop energy density $\rho$ in units of the total energy density $\rho_{\rm tot}$,
\begin{align}
\Omega\left(t\right) & = \frac{\rho\left(t\right)}{\rho_{\rm tot}\left(t\right)} \approx \frac{\mu}{3H^2 M_{\rm Pl}^2} \int_0^\infty d\ell\:\ell\,n_{\rm RM}\left(\ell,t\right) \\
& \approx \frac{\mu}{3H^2 M_{\rm Pl}^2}\frac{2}{y_r^{1/2}t_{\rm ini}^{1/2}}\frac{A_r}{t_{\rm eq}^{3/2}}\left(\frac{a_{\rm eq}}{a\left(t\right)}\right)^3 \,.
\end{align}
Evaluating this expression today, i.e., at $t=t_0$, we find
\begin{equation}
h^2\Omega\left(t_0\right) \simeq 1.1 \times 10^{-13}\left(\frac{G\mu}{10^{-19}}\right)\left(\frac{10^2\,\textrm{s}}{t_{\rm ini}}\right)^{1/2} \,.
\end{equation}
We thus conclude that the present-day contribution from cosmic strings to the total energy budget of the Universe is negligibly small and far away from causing an overclosure problem. Unsurprisingly, it is therefore indeed a viable possibility that cosmic strings formed at a relatively low energy scale in the early Universe, $v \lesssim 10^9\,\textrm{GeV}$, and remained present in the Universe until today.

%%%%%%%%%%%%%%%%%%%%%%%%%%%%%%%%%%%%%%%%%%%%%%%%%%%%%%%%%%%%%%%%%%%%%%%%%%%%%%%%%%%%%%%%%%%%%%%%%%%%

\section{GWB spectrum in the BOS model}
\label{sec:bos}

The sharp drop-off of the GWB spectrum from fundamental loop oscillations at $f_{\rm cut}$ is tied to the assumption that, as soon as the string network has reached the scaling regime, loops are continuously created at the same length scale, i.e., with a length corresponding to a fixed fraction of the particle horizon, $x = \ell/d_h = \alpha \simeq 0.05$. In most discussions of the GWB spectrum from cosmic strings in the literature, this assumption represents a reasonable approximation that simplifies analytical and numerical calculations but has little impact on key features of the spectrum. In this paper, we, however, find ourselves in a different situation. The series of sharp dips in the GWB spectrum from low-scale strings that we observe is at the core of our analysis\,---\,and a direct consequence of the fact that we work with one fixed $x$ value in Eq.~\eqref{eq:fcut} (namely, $x = \alpha$) rather than a distribution of $x$ values. In this section, we shall therefore drop this assumption and generalize our computation of the GWB spectrum to a distribution of initial $x$ values, specifically, the distribution of $x$ values that can be extracted from the numerical simulations carried out by BOS~\cite{Blanco-Pillado:2013qja}. This analysis will enable us to conclude that our results based on the VOS model remain valid as long as the distribution of initial loop lengths continues to be very narrow, which it indeed is in the BOS model.

To study the impact of a smooth distribution of initial loop lengths on the GWB spectrum, we need to return to the computation of the loop number density $n$. In full generality, $n$ can be written as an integral over the loop production function $f$ and an appropriate redshift factor, 
\begin{equation}
\label{eq:nltflt}
n\left(\ell,t\right) = \int_{t_{\rm ini}}^t dt'\: f\left(\ell',t'\right)\left(\frac{a\left(t'\right)}{a\left(t\right)}\right)^3 \,.
\end{equation}
Here, $f\left(\ell,t\right)dl$ counts the number of loops whose length falls into the interval $\left[\ell,\ell+d\ell\right]$ that is created per spatial volume per time interval. In Eq.~\eqref{eq:nltflt}, $\ell$ and $t$ are the independent arguments of the loop number density, the integration variable $t'$ scans over all possible times of loop production, and $\ell'$ is a function in dependence of $\ell$, $t$, and $t'$ that yields the length at time $t'$ of a loop that has length $\ell$ at time $t$. From Eq.~\eqref{eq:dldt}, we immediately find
\begin{equation}
\label{eq:lltt}
\ell' = \ell + \Gamma G\mu\left(t-t'\right) \,.
\end{equation}
In the scaling regime, $f$ factorizes into a time-dependent factor, $1/d_h^5$, and a time-independent factor, $\tilde{f}$, that only depends on the scaling variable $x = \ell/d_h$, 
\begin{equation}
f\left(\ell,t\right) = \frac{\tilde{f}\left(x\right)}{d_h^5} \,, \qquad x = \frac{\ell}{d_h} \,, \qquad d_h = d_h\left(t\right) \,.
\end{equation}
Here, the factor $1/d_h^5$, which expresses the scaling law describing the behavior of $f$ as a function of time, follows directly from dimensional analysis. Meanwhile, the nontrivial information on the distribution of initial $x$ values is encoded in the second factor, i.e., the function $\tilde{f}$.

Let us now focus on loop production during radiation domination. In this case, neglecting again any changes in the number of relativistic degrees of freedom, the particle horizon is given by $d_h = 2t$, and the scale factor $a$ grows like $t^{1/2}$, such that $n$ in Eq.~\eqref{eq:nltflt} can be written as
\begin{equation}
n_{\rm RR}\left(\ell,t\right) = \int_{t_{\rm ini}}^t dt'\: \frac{\tilde{f}\left(x'\right)}{\left(2t'\right)^5}\left(\frac{t'}{t}\right)^{3/2} \,, \qquad x' = \frac{\ell'}{2t'} \,.
\end{equation}
Here, $x'$ and $t'$ are related to each other via [see Eq.~\eqref{eq:lltt}]
\begin{equation}
t' = \frac{\ell + \Gamma G\mu t}{2x' + \Gamma G\mu} \,,
\end{equation}
which allows us to express $n_{\rm RR}$ as an integral over $x'$,
\begin{equation}
n_{\rm RR}\left(\ell,t\right) = \frac{\int_{x'\left(t\right)}^{x'\left(t_{\rm ini}\right)} dx'\:\left(2x'+\Gamma G\mu\right)^{3/2}\tilde{f}\left(x'\right)}{2^4\,t^{3/2}\left(\ell + \Gamma G\mu t\right)^{5/2}} \,,
\end{equation}
where $x'\left(t_{\rm ini}\right)$ and $x'\left(t\right)$ are shorthand notations for
\begin{equation}
x'\left(t'\right) = \frac{\ell + \Gamma G\mu\left(t-t'\right)}{2t'}  \,, \qquad t' = t_{\rm ini},\,t \,.
\end{equation}
As we are only interested in $x$ values of $\mathcal{O}\left(\alpha\right)$ and small string tensions, $\Gamma G\mu \ll 2\alpha$, we are able to neglect the $\Gamma G\mu$ term in the $x'$ integral in the numerator, such that
\begin{align}
\label{eq:nx32f}
n_{\rm RR}\left(\ell,t\right) & \approx \frac{\mathcal{N}\left(t,t_{\rm ini}\right)}{t^{3/2}\left(\ell + \Gamma G\mu t\right)^{5/2}} \,, \\
\mathcal{N}\left(t,t_{\rm ini}\right) & = \frac{1}{2^{5/2}}\,\int_{x'\left(t\right)}^{x'\left(t_{\rm ini}\right)} dx'\:x'^{3/2}\tilde{f}\left(x'\right) \,.
\end{align}
The expression in Eq.~\eqref{eq:nx32f} marks the final expression for $n_{\rm RR}$ that we are able to derive without making any further assumptions regarding the loop production function.

\subsection{Single initial loop length}

Before we turn to the BOS distribution of initial $x$ values, let us first review the evaluation of Eq.~\eqref{eq:nx32f} in the limit of a single initial loop length. To this end, we write the normalization factor in the numerator as
\begin{equation}
\label{eq:Norm}
\mathcal{N}\left(t,t_{\rm ini}\right) = \int_{\ln x'\left(t\right)}^{\ln x'\left(t_{\rm ini}\right)} d\left(\ln x'\right)\:\rho\left(\ln x'\right) \,,
\end{equation}
where $\rho$ is the measure for the distribution of initial $x$ values that is numerically determined by BOS in Ref.~\cite{Blanco-Pillado:2013qja},%
\footnote{In Ref.~\cite{Blanco-Pillado:2013qja}, $x^{5/2}\tilde{f}\left(x\right)$ is denoted by $\alpha^{5/2}f_r\left(\alpha\right)$.}
\begin{equation}
\rho\left(\ln x\right) = \left[\frac{1}{2^{5/2}}\,x^{5/2}\tilde{f}\left(x\right)\right]_{x = e^{\ln x}} \,,
\end{equation}
and choose a Dirac delta distribution ansatz for $\rho$, 
\begin{equation}
\label{eq:deltaansatz}
\rho\left(\ln x\right) = A_r\,\delta\left(\ln\left(x/\alpha\right)\right) \,,
\end{equation}
where, as before, $A_r = \mathcal{F}\,C_r\,y_r^{1/2}\simeq 0.17$. At the level of the loop production function, this ansatz corresponds to 
\begin{equation}
f\left(\ell,t\right) = \frac{\mathcal{F}\,C_r}{2\alpha\,t^4}\,\delta\left(\ell-2\alpha\,t\right) \,,
\end{equation}
where we used $y_r = 2\alpha$. By construction, we thus retrieve the standard loop production function in the VOS model in the limit of a single initial loop length. At the same time, it is now trivial to evaluate the integral in Eq.~\eqref{eq:Norm},
\begin{equation}
\label{eq:Nstandard}
\mathcal{N}\left(t,t_{\rm ini}\right) \quad\overset{\eqref{eq:deltaansatz}}{\rightarrow}\quad A_r\,\Theta\left(t-t_*\right)\Theta\left(t_*-t_{\rm ini}\right) \,,
\end{equation}
where $t_*$ is given by Eq.~\eqref{eq:tstar} after identifying $y_*$ with $y_r$. Combining this result with Eq.~\eqref{eq:nx32f}, we finally obtain the standard result for the RR loop number density in the VOS model in the limit of a single initial loop length,
\begin{equation}
\label{eq:nstandard}
n_{\rm RR}\left(\ell,t\right) \approx \frac{A_r\,\Theta\left(t-t_*\right)\Theta\left(t_*-t_{\rm ini}\right)}{t^{3/2}\left(\ell + \Gamma G\mu t\right)^{5/2}} \,.
\end{equation}

\subsection{Log-normal distribution of initial loop lengths}

After reviewing the usual limit of a single initial loop length in the previous section, we are now ready to generalize our results to a broader distribution of initial loop lengths. To do so, we note that the numerical results for the density function $\rho$ obtained by BOS in Ref.~\cite{Blanco-Pillado:2013qja} (see the left panel of Fig.~2 in this paper) can be approximately described by a Gaussian with mean $\nu = \ln\alpha \simeq -3.0$ and standard deviation $\sigma \simeq 0.14$,
\begin{equation}
\label{eq:rhoBOS}
\rho\left(\ln x\right) \simeq \frac{A_r}{\sqrt{2\pi}\,\sigma}\,e^{-\frac{\left(\ln x - \nu\right)^2}{2\sigma^2}} \,,
\end{equation}
which corresponds to a loop production function of
\begin{equation}
f\left(\ell,t\right) = \frac{A_r}{\sqrt{2\pi}\,\sigma\,\ell^{5/2}t^{5/2}}\,e^{-\frac{\left(\ln\left(\ell/\left(2t\right)\right) - \nu\right)^2}{2\sigma^2}} \,.
\end{equation}
The density function $\rho$ in Eq.~\eqref{eq:rhoBOS} strikes a balance: on the one hand, it successfully captures the nonzero width of the exact numerical BOS distribution close to its peak; on the other hand, it is still simple enough to allow for a purely analytical computation of the RR loop number density. Indeed, working with the simple $\rho$ in Eq.~\eqref{eq:rhoBOS}, the integral in Eq.~\eqref{eq:Norm} can again be solved exactly,
\begin{equation}
\mathcal{N}\left(t,t_{\rm ini}\right) \quad\overset{\eqref{eq:rhoBOS}}{\rightarrow}\quad \frac{A_r}{2}\left.\textrm{erf}\left(\frac{\ln x'-\nu}{\sqrt{2}\sigma}\right)\right|_{\ln x'\left(t\right)}^{\ln x'\left(t_{\rm ini}\right)} \,,
\end{equation}
where $\textrm{erf}\left(z\right)$ denotes the Gauss error function. It is straightforward to show that this expression for $\mathcal{N}$ reproduces the standard result in Eq.~\eqref{eq:Nstandard} in the limit $\sigma \rightarrow 0$. Together with Eq.~\eqref{eq:nx32f}, we thus arrive at the following generalized result for the RR loop number density, which properly accounts for the finite-width distribution of initial $x$ values in the BOS numerical simulation,
\begin{equation}
\label{eq:nRRBOS}
\boxed{n_{\rm RR}\left(\ell,t\right) \approx \frac{A_r\left(\textrm{erf}_{t_{\rm ini}}-\textrm{erf}_t\right)/2}{t^{3/2}\left(\ell + \Gamma G\mu t\right)^{5/2}} \,,}
\end{equation}
where $\textrm{erf}_{t_{\rm ini}}$ and $\textrm{erf}_t$ are shorthand notations for
\begin{equation}
\textrm{erf}_{t'} = \textrm{erf}\left(\frac{\ln x'\left(t'\right)-\nu}{\sqrt{2}\sigma}\right) \,, \qquad t' = t_{\rm ini},\, t \,.
\end{equation}
To the best of our knowledge, a result of this form has not been derived before, even though it constitutes a more accurate approximation of the exact numerical BOS result than the standard RR loop number density in Eq.~\eqref{eq:nstandard}.

In view of this result, two comments are in order. First, we stress that our analysis is based on the approximate log-normal distribution in Eq.~\eqref{eq:rhoBOS}, which is motivated by the fact that we are primarily interested in analytical expressions that are parametrically related to the standard case of a single initial loop length (i.e, the case of a single initial loop length can be retrieved in the limit $\sigma \rightarrow 0$). Still, the formalism outlined in this section is fully general and could also be used for a purely numerical evaluation of the normalization factor $\mathcal{N}$ in Eq.~\eqref{eq:Norm}, e.g., based on the exact numerical BOS results for $\rho$ in Ref.~\cite{Blanco-Pillado:2013qja} or any future refinement thereof. However, instead of carrying out such a numerical analysis, we will restrict ourselves in the following to the expressions stated above\,---\,which we will use shortly to scan over a range of density functions (by varying the parameter $\sigma$) and hence draw a bigger picture of the uncertainties inherent to our analysis.

Our second comment is that the applicability of the RR loop number density in Eq.~\eqref{eq:nRRBOS} extends beyond the discussion in this paper. We argue that, in any analysis relying on the BOS loop number densities, it would in principle be more accurate to work with Eq.~\eqref{eq:nRRBOS} than with Eq.~\eqref{eq:nstandard}. In practice, though, noticeable numerical effects will only occur in the case of low-scale strings. We explicitly checked that, for high-scale strings, all corrections to the GWB spectrum arising from our refined loop number densities remain at the sub-percent level.

For our purposes, we actually require the RM loop number density, which can, however, be directly obtained from the RR loop number density. Indeed, the RM loop number density simply follows from evaluating $n_{\rm RR}$ at time $t_{\rm eq}$ and loop length $\ell_{\rm eq} = \ell + \Gamma G\mu\left(t-t_{\rm eq}\right)$ (i.e., the length that a loop with length $\ell$ at time $t$ had previously at time $t_{\rm eq}$), in combination with a redshift factor,
\begin{equation}
n_{\rm RM}\left(\ell,t\right) = n_{\rm RR}\left(\ell_{\rm eq},t_{\rm eq}\right) \left(\frac{a_{\rm eq}}{a\left(t\right)}\right)^3\,.
\end{equation}
Then, using that $\ell_{\rm eq} + \Gamma G\mu\,t_{\rm eq}= \ell + \Gamma G\mu\,t$,  we find
\begin{equation}
\label{eq:nRMBOS}
n_{\rm RM}\left(\ell,t\right) \approx \frac{A_r\left(\textrm{erf}_{t_{\rm ini}}-\textrm{erf}_{t_{\rm eq}}\right)/2}{t_{\rm eq}^{3/2}\left(\ell + \Gamma G\mu t\right)^{5/2}}\left(\frac{a_{\rm eq}}{a\left(t\right)}\right)^3 \,,
\end{equation}
where $\textrm{erf}_{t_{\rm eq}}$ corresponds to $\textrm{erf}_t$ at $t = t_{\rm eq}$. Again, this result represents the numerical BOS result more accurately than the standard RM loop number density in Eq.~\eqref{eq:nRM2}.

Finally, let us estimate the ratio of values of the scale factor in Eq.~\eqref{eq:nRMBOS}. Assuming perfect radiation domination all the way down to $t=t_{\rm eq}$, the value of the scale factor at matter--radiation equality can be written as
\begin{equation}
a_{\rm eq} = a_0\left(2\,\Omega_{\rm r}^{1/2}H_0\,t_{\rm eq}\right)^{1/2} \,, \quad h^2\Omega_{\rm r} \simeq 4.2 \times 10^{-5} \,.
\end{equation}
Meanwhile, assuming perfect matter domination at all times $t>t_{\rm eq}$, we can use Eq.~\eqref{eq:atm} for $a\left(t\right)$, such that
\begin{equation}
\label{eq:nRMBOS2}
n_{\rm RM}\left(\ell,t\right) \approx R\,\frac{A_r\left(\textrm{erf}_{t_{\rm ini}}-\textrm{erf}_{t_{\rm eq}}\right)/2}{t^2\left(\ell + \Gamma G\mu t\right)^{5/2}} \,,
\end{equation}
where $R$ takes care of all remaining prefactors,
\begin{equation}
R = \frac{\left(2\,\Omega_{\rm r}^{1/2}H_0\right)^{3/2}}{\left(\sfrac{3}{2}\,\Omega_{\rm m}^{1/2}H_0\right)^2} \,.
\end{equation}

The treatment of the ratio $a_{\rm eq}/a$ leading from Eq.~\eqref{eq:nRMBOS} to Eq.~\eqref{eq:nRMBOS2} is rather rough and fails to account for the smooth transition from radiation domination to matter domination. As a consequence, the numerical factor $R$ in Eq.~\eqref{eq:nRMBOS2} slightly shifts the normalization of the loop number density and hence the resulting GWB spectrum. To assess the size of this effect, we compute the GWB spectrum based on Eq.~\eqref{eq:nRMBOS2} in the limit $\sigma \rightarrow 0$ and compare its normalization to the normalization of the standard GWB spectrum based on the loop number density in Eq.~\eqref{eq:nloop} and a fully numerical evaluation of the scale factor. This comparison shows that the naive factor $R$ is off by roughly a factor $1.5$, which we can easily compensate for by replacing $R$ in Eq.~\eqref{eq:nRMBOS2}  by $R'=R/1.5$,
\begin{equation}
\label{eq:nRMBOS3}
\boxed{n_{\rm RM}\left(\ell,t\right) \approx R'\,\frac{A_r\left(\textrm{erf}_{t_{\rm ini}}-\textrm{erf}_{t_{\rm eq}}\right)/2}{t^2\left(\ell + \Gamma G\mu t\right)^{5/2}}} \,.
\end{equation}
This expression represents our final result for the RM loop number density in the BOS model, which accounts for the log-normal distribution of initial $x$ values. It is instructive to compare this result for $n_{\rm RM}$ to the result for $n_{\rm RR}$ in Eq.~\eqref{eq:nRRBOS}. Both expressions exhibit a similar structure, but differ in terms of the power-law exponents in the denominator ($t^{3/2}$ versus $t^2$), the arguments of the error function in the numerator ($\textrm{erf}_t$ versus $\textrm{erf}_{t_{\rm eq}}$), and the overall amplitude factor ($A_r$ versus $R' A_r$). Here, the new factor $R'$ can in particular be thought of as the late-time limit of a transition function that accounts for the change of the scale factor across the transition from radiation domination to matter domination.

\begin{figure}
\begin{center}
\includegraphics[width = 0.48\textwidth]{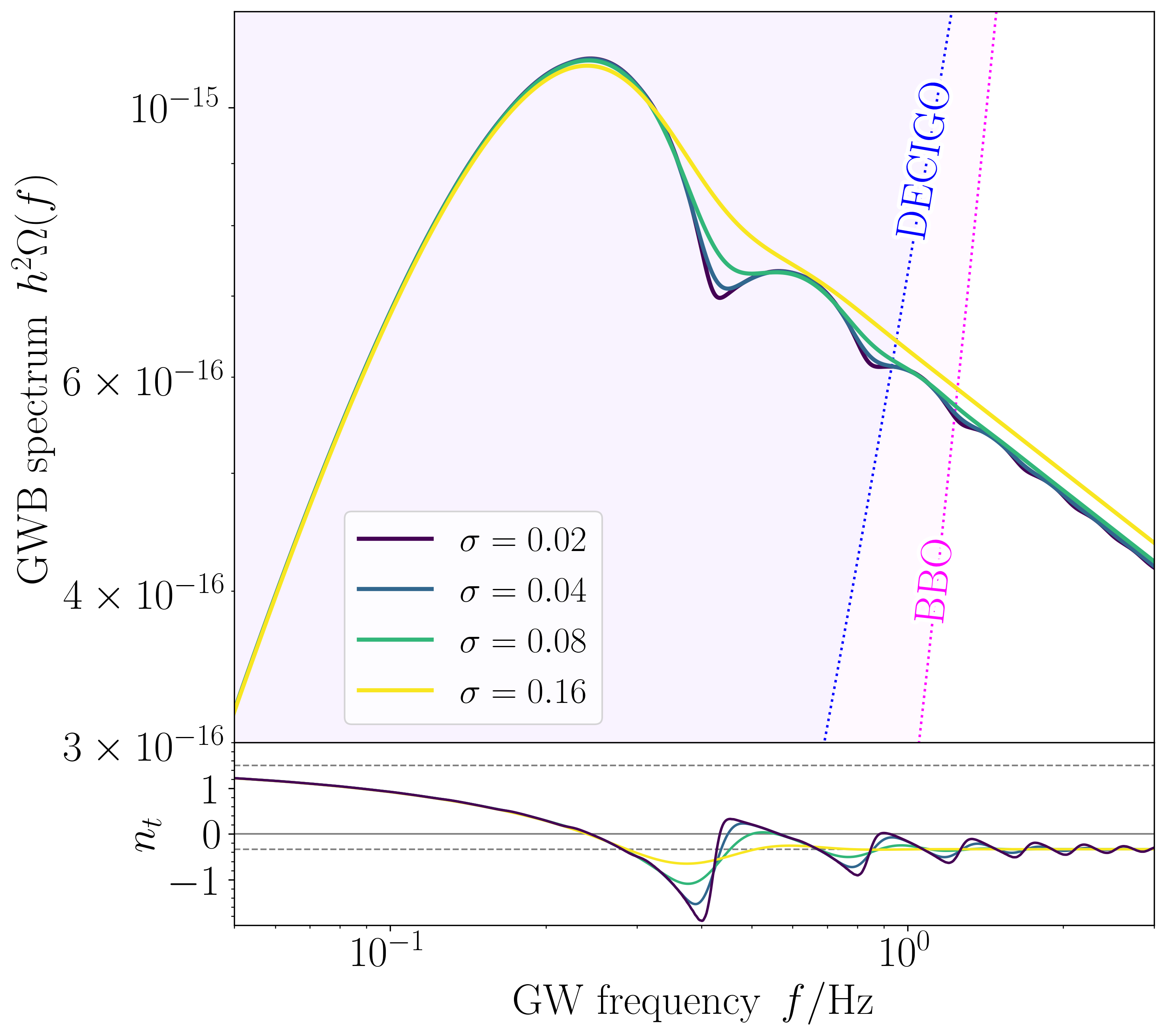}
\end{center}
\caption{GWB spectrum for benchmark point 2 based on the generalized BOS loop number density $n_{\rm RM}$ in Eq.~\eqref{eq:nRMBOS3}, which assumes a log-normal distribution of the initial loop length at the time of formation. The mean of the distribution is fixed at $\nu = -3.0$, while its standard variation is varied, $\sigma \in \left\{0.02, 0.04, 0.08, 0.16\right\}$. The lower panel shows the spectral index $n_t$ as a function of frequency; see Eq.~\eqref{eq:nt}. The horizontal dashed lines in this panel indicate the asymptotic values of the spectral index: $n_t \rightarrow \sfrac{3}{2}$ at low frequencies and $n_t \rightarrow -\sfrac{1}{3}$ at high frequencies; see the GWB spectra in Fig.~\ref{fig:4spectra}.}
\label{fig:alphadistr}
\end{figure}

In the last step of our analysis, we use the loop number density in Eq.~\eqref{eq:nRMBOS3} to compute the resulting GWB spectrum for a distribution of initial $x$ values with mean $\nu = -3.0$ and standard deviation $\sigma \in \left\{0.02, 0.04, 0.08, 0.16\right\}$; see Fig.~\ref{fig:alphadistr}. The range spanned by these values includes the value that we read off from Fig.~2 in Ref.~\cite{Blanco-Pillado:2013qja}, $\sigma \simeq 0.14$, and illustrates at the same time the range of possibilities that we may expect in different models. As evident from Fig.~\ref{fig:alphadistr}, broader distributions of initial loop lengths tend to wash out the series of peaks and dips in the spectrum more strongly. However, even for $\sigma$ values as large as $\sigma = 0.16$, the oscillatory modulation of the $f^{-1/3}$ power-law behavior partially survives and could be detected by feature GW experiments. To highlight this point, we show in the lower panel of Fig.~\ref{fig:alphadistr} the index of the GWB spectrum as a function of frequency,
\begin{equation}
\label{eq:nt}
n_t = \frac{d\ln h^2\Omega_{\rm GW}\left(f\right)}{d\ln f} \,,
\end{equation}
which displays prominent oscillations around $n_t = -1/3$ at frequencies $f \gtrsim f_{\rm cut}$. This is even true for $\sigma = 0.16$, in which case BBO and DECIGO may be able to detect at least one local minimum and one local maximum in $n_t$. In summary, we conclude that, as long as the distribution of initial loop lengths remains sufficiently narrow, the mean features of the GWB spectrum that we derived in the VOS model in the limit of a single initial loop length have a realistic chance of surviving in the BOS model.

%%%%%%%%%%%%%%%%%%%%%%%%%%%%%%%%%%%%%%%%%%%%%%%%%%%%%%%%%%%%%%%%%%%%%%%%%%%%%%%%%%%%%%%%%%%%%%%%%%%%

\section{Conclusions}
\label{sec:conclusions}

In this paper, we studied the GWB spectrum from local field-theory strings associated with an underlying symmetry breaking scale in the range from $v \sim 10^2\,\textrm{GeV}$ to $v \sim 10^9\,\textrm{GeV}$. Strings of this type, which we dub ``low-scale strings'', possess a very small tension, in the range from $G\mu \sim 10^{-33}$ to $G\mu \sim 10^{-19}$, and thus lose energy via the emission of GWs only very slowly. Consequently, if loops are born with sufficiently large lengths in the early Universe, one arrives at the conclusion that no string loop has had enough time yet to fully evaporate\,---\,all string loops created in the early Universe still exist today, and none of them has yet reached zero length.

The shortest loops today correspond to the population of loops created in the early Universe at $t_{\rm ini}$, where $t_{\rm ini}$ is the earliest time when loop production in the network is no longer suppressed by thermal friction and GW emission from loops is no longer subdominant to particle emission from loops. According to this definition, $t_{\rm ini}$ marks the time from which on we can rely on the standard computation of the GWB signal from strings,
\begin{equation}
\Omega_{\rm GW}\left(f\right) = \frac{16\pi}{3H_0^2}\left(G\mu\right)^2 \sum_{k=1}^{k_{\rm max}}\frac{k P_k}{f}\,\int_{t_{\rm ini}}^{t_0}dt \left[\cdots\right] \,.
\end{equation}
There are several well-motivated choices for $t_{\rm ini}$, which we discussed in detail in Sec.~\ref{subsec:lini}. In particular, we find that $t_{\rm ini}$ can become exceptionally large in the case of a small string tension. As a result, the evolution of a network of low-scale strings is characterized by a large hierarchy between the temperature scale at which the network forms, $T_{\rm form}$, and the scale at which the standard picture of a scaling network losing energy via GW emission ($T_{\rm fric}$ or $T_{\rm cusp}$) becomes applicable; see Fig.~\ref{fig:Tscales}.

Our main result in this paper consists in the observation that combining (i) a small string tension $G\mu$ with (ii) a large initial time $t_{\rm ini}$ prevents loops from fully evaporating, which in turn has important implications for the GWB spectrum: at the level of the GWB spectrum from fundamental loop oscillations, we find a sharp cutoff frequency $f_{\rm cut}$ [see Eq.~\eqref{eq:fcut}], which translates to a series of peaks and dips at integer multiples of $f_{\rm cut}$ in the GWB spectrum from all oscillation modes. We thus arrive at a GWB spectrum with distinct features, which, remarkably enough, may be within the sensitivity reach of upcoming GW experiments. There is especially a sweet spot in parameter space around $G\mu \sim 10^{-19}$ and $T_{\rm ini} \sim 100\,\textrm{keV}$ that will be probed by BBO and DECIGO.

In this paper, we investigated the GWB spectrum from low-scale strings based on the VOS and BOS models both analytically and numerically. In particular, we refined the well-known BOS loop number densities $n_{\rm RR}$ and $n_{\rm RM}$, taking into account the finite-width distribution of initial loop lengths observed in the BOS numerical simulations; see Eqs.~\eqref{eq:nRRBOS} and \eqref{eq:nRMBOS}. As a result of this analysis, we found that a finite-width distribution of initial loop lengths can wash out the peaks and dips in the GWB spectrum. For a sufficiently narrow distribution, these features, however, partially survive and thus remain an interesting observational target; see Fig.~\ref{fig:alphadistr}. Besides that, our generalized results for $n_{\rm RR}$ and $n_{\rm RM}$ can be used in any future analysis of cosmic strings that intends to work with the BOS loop number densities.

Both the VOS model and the BOS model apply to scaling string networks. The scaling behavior of the network, however, becomes violated in models where particle emission begins to dominate over GW emission as soon as a loop has shrunk below a certain critical size. A nonscaling model of this type has been discussed in Ref.~\cite{Auclair:2019jip}. Using the nonscaling loop number densities worked out in this paper, we already performed a preliminary computation of the GWB spectrum, which led us to nearly identical results as those obtained for the VOS and BOS models. These findings, which we will present in more detail elsewhere, confirm the robustness of the results in the present paper. We thus conclude that strings with a small tension deserve more attention, especially, strings created at an energy scale in the vicinity of $v \sim 10^9\,\textrm{GeV}$, which, as we demonstrated in this paper, can lead to an observable GWB spectrum characterized by distinct features. String formation at an energy scale of around $v \sim 10^9\,\textrm{GeV}$ can occur in many models of physics beyond the Standard Model. In future work, it will be interesting to identify the most promising microscopic models that can realize the GWB spectrum discussed in this paper.

%%%%%%%%%%%%%%%%%%%%%%%%%%%%%%%%%%%%%%%%%%%%%%%%%%%%%%%%%%%%%%%%%%%%%%%%%%%%%%%%%%%%%%%%%%%%%%%%%%%%

\section{Acknowledgments}

The authors would like to thank Jos\'e Juan Blanco-Pillado, Ken Olum, Pedro Schwaller, and Qaisar Shafi for fruitful discussions and comments. The work of K.\,S.\ and T.\,S.\ is supported by Deutsche Forschungsgemeinschaft (DFG) through the Research Training Group (Graduiertenkolleg) 2149: Strong and Weak Interactions\,---\,from Hadrons to Dark Matter.

%%%%%%%%%%%%%%%%%%%%%%%%%%%%%%%%%%%%%%%%%%%%%%%%%%%%%%%%%%%%%%%%%%%%%%%%%%%%%%%%%%%%%%%%%%%%%%%%%%%%

\bibliography{manuscript_2}

%%%%%%%%%%%%%%%%%%%%%%%%%%%%%%%%%%%%%%%%%%%%%%%%%%%%%%%%%%%%%%%%%%%%%%%%%%%%%%%%%%%%%%%%%%%%%%%%%%%%

\end{document}